\documentclass[journal]{IEEEtran}

\usepackage{amsmath,amsfonts,amssymb}
\usepackage{algorithmic}
\usepackage{algorithm}
\usepackage{array}
\usepackage[caption=false,font=normalsize,labelfont=sf,textfont=sf]{subfig}
\usepackage{textcomp}
\usepackage{stfloats}        
\usepackage{url}
\usepackage{verbatim}
\usepackage{graphicx}
\usepackage{cite}
\usepackage{booktabs}
\usepackage{multirow}
\usepackage{adjustbox}
\usepackage{xcolor}
\usepackage{xspace}          
\usepackage{tikz}
\usepackage{hyperref}
\hypersetup{colorlinks=true,linkcolor=blue,citecolor=blue,urlcolor=blue}

\graphicspath{{figures/}}
\usepackage{orcidlink}

\newcommand{\sysname}{\textsc{FailBench}\xspace}

\newcommand{\markm}{\textit{m}}
\newcommand{\marka}{\textcolor{gray!85}{\textit{a}}}
\newcommand{\vala}[1]{\textcolor{gray!85}{#1}~\marka}

\begin{document}

\title{\sysname: Evaluating Fault Tolerance Across Distributed Training Architectures}

\author{Khaled~Aljbab\,\orcidlink{0009-0003-4491-2191}
and Amine~Barrak\,\orcidlink{0000-0002-0046-2454}
\thanks{K. Aljbab and A. Barrak are with the Department of Computer Science
and Engineering, Oakland University, Rochester, MI 48309 USA.
Corresponding author: A. Barrak (e-mail: aminebarrak@gmail.com).}%
}

\markboth{}%
{Aljbab \MakeLowercase{\textit{and}} Barrak: \sysname: Cross-Architecture Fault Tolerance for Distributed DL}

\maketitle

\begin{abstract}
Distributed deep learning relies on data, pipeline, tensor, and
hybrid parallelism, yet fault-tolerance mechanisms are typically
evaluated only on the architecture for which they were designed.
This leaves practitioners with little guidance when choosing
mechanisms across architectures. \sysname{} provides a unified
evaluation harness covering seven distributed training
architectures, eight crash-fault-tolerance mechanisms and a
no-FT baseline, and single, concurrent, and cascading fail-stop
failures. We evaluate 142 (architecture, mechanism, trace)
combinations on an 8$\times$V100 cluster, with per-rank checkpoint
states ranging from 205\,MB to 2.7\,GB.

Three findings emerge. First, no mechanism is universally best:
on A2, disk checkpointing has the lowest steady-state overhead
(0.5\%), in-memory replication restores fastest
($\sim$17\,ms), and just-in-time checkpointing avoids periodic
steady-state checkpoint cost but incurs $\sim$0.9\,s upon
failure. When process-group re-formation takes seconds,
mechanisms differ more in runtime overhead than restore speed.
Second, gossip training increases sample throughput by 17.9\%
after losing a worker, yet shows no detectable improvement in
loss progress over a matched no-fault baseline. Third, mechanism
cost depends strongly on architecture: in-memory replication
overhead ranges from 3.7\% to 176\%. We translate these findings
into a decision framework for selecting fault-tolerance mechanisms
and release \sysname{} as an open artifact.
\end{abstract}


\section{Introduction}
\label{sec:introduction}
\IEEEPARstart{T}{raining} large neural networks has become an exercise in
endurance computing. Modern language and vision models have grown from
millions to hundreds of billions of parameters in under a decade
\cite{zhang2022opt,llama3card}, and neither the model nor the dataset
fits on a single accelerator any longer.
Practitioners therefore rely on \emph{distributed} training: the model,
the data, or both are partitioned across many GPUs that exchange gradients
or activations on every training step. The dominant parallelism
strategies (data, pipeline, and tensor parallelism, often composed
into three-dimensional
hybrids~\cite{shoeybi2019megatron,rasley2020deepspeed}) each carry
their own communication pattern, memory layout, and failure topology.
Frontier-scale runs span weeks of wall time on thousands of
GPUs~\cite{jiang2024megascale}, consume megawatts of power, and cost
millions of dollars per training cycle.

At this scale, failure is routine rather than exceptional. Meta's
OPT-175B log reports roughly one hundred hardware failures over a
two-month run~\cite{zhang2022opt}; the Llama-3 405B campaign was
interrupted more than 400 times in 54 days, dominated by GPU and HBM
failures~\cite{llama3card}; and reports from
ByteDance~\cite{jiang2024megascale} and Alibaba~\cite{wu2024falcon}
document crashes, link drops, fail-slow
stragglers, and silent data corruption.

In response, the systems and machine learning communities have
produced a collection of fault tolerance (FT) mechanisms,
cataloged by several
surveys~\cite{cappello2014toward,shi2020fault,khan2024hierarchical}.
The common approach periodically saves model and optimizer state
and replays from the latest checkpoint after a failure; recent work
accelerates it through in-memory replication~\cite{wang2023gemini},
asynchronous and lazy disk
writes~\cite{maurya2024datastates,wan2024bytecheckpoint}, and
just-in-time capture triggered by an impending
preemption~\cite{gupta2024jit}. A second family duplicates the work,
running each pipeline stage on more than one rank so a
surviving peer takes over
immediately~\cite{thorpe2023bamboo,jang2023oobleck,athlur2022varuna}.
A third applies coding theory, distributing redundant computation so
that any sufficient subset of workers reconstructs the aggregate
update~\cite{tandon2017gradient}. A fourth accepts convergence
cost in exchange for graceful degradation: decentralized parallel
SGD~\cite{lian2017can}, PyTorch Elastic~\cite{pytorch_elastic}, and
Horovod Elastic~\cite{sergeev2018horovod} keep training alive at a
smaller cluster size instead of crashing.

Despite this diversity, the literature shares a methodological
limitation: \emph{each mechanism is evaluated only inside its own
design space}. Checkpoint systems are compared against other
checkpoint variants~\cite{wang2023gemini,maurya2024datastates},
redundancy systems against vanilla pipeline
parallelism~\cite{thorpe2023bamboo,jang2023oobleck}, and gossip-SGD
inside the peer-to-peer overlay~\cite{lian2017can,koloskova2020unified}.
No published study measures these mechanisms head-to-head across the
spectrum of training architectures on the same cluster under a common
fault schedule.
A practitioner who has chosen an architecture for performance reasons
therefore has no quantitative basis for picking the FT strategy that
minimizes expected wall time given the cluster's mean time between
failures.

We propose \sysname{}, a harness for evaluating crash-fault tolerance
across distributed training architectures under controlled fault
schedules. It covers 133 of the 171 viable (architecture, mechanism,
trace) tuples plus all nine structurally infeasible tuples. Four
conditions bound what follows. We inject fail-stop crashes only, so
Byzantine behavior, silent data corruption, and fail-slow stragglers
are out of scope. The mechanisms span checkpointing, state replication,
and reconfiguration, but not coded computation, which needs an
aggregation topology we do not implement. The schedule is fixed
across architectures except for injection time, which scales with
per-rank state size so reactive mechanisms can finish an emergency
save before the crash. We measure at 8 GPUs on
$0.2$--$2.7$\,GB of per-rank checkpoint state, parameters plus
optimizer buffers, so the transferable results are structural:
which architectures halt, which pairings are infeasible, and the
shape of the overhead-versus-recovery trade-off, rather than absolute
overheads at frontier scale.

\smallskip\noindent\textbf{Contributions.} This paper makes four
contributions.
\begin{enumerate}
\item \sysname, a unified harness that pairs seven distributed
training architectures with eight crash-fault tolerance
mechanisms and a no-FT baseline
under a shared fault injector, enabling the first cross-cutting
comparison on identical hardware and a common trace family.

\item A compatibility map of 63 architecture--mechanism pairs
that distinguishes viable, degenerate, and \emph{structurally
infeasible} configurations, including three infeasible pairs
yielding nine tuples across the trace family.

\item A 142-tuple empirical study showing that (i) six
configurations halt completely on a single worker crash, while
A1-async and A6 gossip degrade gracefully and gossip's raw
throughput \emph{rises} post-fault without improving learning
efficiency, and (ii) disk, in-memory, and JIT checkpointing form a
Pareto frontier ranging from zero periodic checkpoint work to
$3.8\%$ measured steady-state overhead, with $17$--$900$\,ms restore
time, on which no mechanism dominates.

\item A measurement-grounded decision framework that translates the
empirical results into principled fault-tolerance selection, minimizing
expected job completion time under user-supplied failure rate, state
size, checkpoint interval, and service-level constraints.
\end{enumerate}
We release \sysname, all configurations, fault schedules, and the
harvesting and analysis scripts, together with the rollup outputs
behind every table, as an open
artifact.\footnote{\url{https://zenodo.org/records/23025990}}
\section{Background}
\label{sec:background}
This section summarizes distributed training architectures and
fault-tolerance mechanisms considered in \sysname{}.

\subsection{Distributed Training Architectures}
\label{sec:bg:arch}

Modern DDL parallelizes SGD through \emph{data},
\emph{pipeline}, and \emph{tensor} parallelism. \sysname{}
studies seven resulting architectures that differ in
communication pattern and failure exposure.

\textbf{(A1) Parameter Server, synchronous and asynchronous.}
Workers compute local gradients and send them to a parameter
server, which applies the optimizer update and distributes
weights. Rank 0 hosts the optimizer and ranks $1,\ldots,N{-}1$
are workers. The synchronous variant uses
\texttt{torch.distributed.gather} and \texttt{broadcast} over
NCCL and waits for all workers. The asynchronous variant uses
per-worker communication over gloo/TCP and applies updates
without waiting for every worker. Both are built on
\texttt{torch.distributed}, without ps-lite or BytePS. Failure
impact therefore depends both on whether the failed rank is the
server or a worker and on the synchronization policy.

\textbf{(A2) Ring All-Reduce.}
Workers form a ring and exchange gradient shards through a
bandwidth-efficient collective. We use PyTorch DDP over NCCL 2.20.
Because all ranks participate in every collective, a single worker
failure stalls training until the ring is repaired.

\textbf{(A3) Hierarchical / Tree All-Reduce.}
We use the same PyTorch DDP/NCCL compute path as A2 but force
\texttt{NCCL\_ALGO=Tree}. The resulting logical tree reduces gradients
upward and broadcasts them downward. Unlike a ring, failure impact
depends on tree position, with internal-node loss generally more
disruptive than leaf loss.

\textbf{(A4) Pipeline Parallelism (PP).}
The model is partitioned by layer depth across workers, while
micro-batches flow through the resulting pipeline. Our implementation
uses \texttt{torch.distributed} point-to-point communication:
activations move forward through \texttt{dist.send}/\texttt{dist.recv}
and gradients backward through the same primitives. Each rank hosts
one stage of a custom feed-forward chain rather than ResNet-50 because
variable-shape convolutional activations do not pipeline cleanly.
Losing any stage halts the pipeline.

\textbf{(A5) Tensor Parallelism (TP).}
Individual layers are split across workers using a hand-rolled
\texttt{torch.distributed} implementation of the Megatron
transposed-MLP pattern: a column-parallel linear layer, GELU, and a
row-parallel linear layer whose partial outputs are combined with
\texttt{all\_reduce}. We use a four-block transposed-MLP stack.
Every rank in the TP group participates in each step, so losing one
rank breaks the distributed matrix operation.

\textbf{(A6) Decentralized Peer-to-Peer (D-PSGD).}
Workers maintain local models and average parameters with neighbors
over a gossip topology~\cite{lian2017can}. We implement D-PSGD using
raw \texttt{torch.distributed} with per-edge Gloo subgroups. Each rank
performs a local SGD update and exchanges post-update parameters with
live ring neighbors under a per-round timeout, averaging parameters
that arrive in time. Timed-out neighbors are removed from the live set
and the overlay reconfigures without a central coordinator. The lack
of a global collective enables graceful degradation after worker loss.

\textbf{(A7) 3D Hybrid (DP$\times$PP$\times$TP).}
Large-model training commonly combines data, pipeline, and tensor
parallelism. We compose A4's pipeline implementation, A5's tensor
parallelism, and a data-parallel \texttt{all\_reduce} across replicas.
At world size 8, the configuration is
$\text{DP}\!\times\!\text{PP}\!\times\!\text{TP}=2\!\times\!2\!\times\!2$.
Its failure topology therefore combines all three axes, and recovery
cost depends on the axis containing the failed worker.

\subsection{Fault Tolerance Mechanisms}
\label{sec:bg:ft}
\label{sec:bg:ft-mechanisms}

We study eight crash-fault-tolerance mechanisms, plus a no-FT
baseline (FT9), spanning checkpointing,
state replication, and runtime reconfiguration.
Table~\ref{tab:ft-summary} summarizes them.

\begin{table}[h]
\caption{Fault-tolerance configurations studied by \sysname.}
\label{tab:ft-summary}
\centering
\small
\setlength{\tabcolsep}{4pt}
\begin{tabular}{@{}l p{0.50\columnwidth} p{0.34\columnwidth}@{}}
\toprule
ID & Mechanism & Family \\
\midrule
FT1  & Periodic disk checkpoint        & Checkpoint \\
FT2  & In-memory replication           & Checkpoint \\
FT3  & Asynchronous / lazy checkpoint  & Checkpoint \\
FT4  & Just-in-time checkpoint         & Checkpoint \\
FT5  & Continuous state replication    & State replication \\
FT6  & $k$-replica state redundancy    & State replication \\
FT7  & Data redistribution             & Reconfiguration \\
FT8  & Elastic process group           & Reconfiguration \\
FT9  & No-FT baseline              & Baseline \\
\bottomrule
\end{tabular}
\end{table}

\subsubsection*{Checkpoint-class recovery (FT1--FT4)}

\textbf{(FT1) Periodic disk checkpoint.}
Every $T$ steps, model, optimizer, and RNG state are serialized to a
shared file system and restored after failure. Persistence cost scales
with model size, while lost work is bounded by $T$ steps; Young--Daly
theory gives the optimal interval under exponential failure
arrivals~\cite{young1974first,daly2006higher}.

\textbf{(FT2) In-memory replication.}
The latest checkpoint is mirrored to a peer worker's GPU or host
memory~\cite{wang2023gemini}, avoiding file-system access during
recovery at the cost of peer memory and communication.

\textbf{(FT3) Asynchronous / lazy disk checkpoint.}
State is staged in a CPU-side buffer and persisted by a background
thread, removing the synchronous disk flush from the training
step~\cite{maurya2024datastates,wan2024bytecheckpoint}. This allows
shorter effective checkpoint intervals at comparable throughput cost.

\textbf{(FT4) Just-in-time checkpoint.}
State is captured only when a preemption signal arrives
~\cite{gupta2024jit}, requiring no periodic steady-state checkpoint work.
Our implementation captures the lead-time-reactive behavior of
Bamboo/Gemini-JIT but not shadow-rank lineage replay. It therefore
incurs emergency state-transfer cost and cannot protect against
unannounced crashes.

\subsubsection*{State replication (FT5--FT6)}

\textbf{(FT5) Continuous state replication.}
The full model and optimizer state are mirrored to a peer at every
training step. FT5 serves as a replication-cost proxy for
compute-redundant systems such as Bamboo and
Oobleck~\cite{thorpe2023bamboo,jang2023oobleck,athlur2022varuna}:
it captures their steady-state bandwidth and memory costs without
duplicating shadow-rank computation. Consequently, it is faithful on
steady-state overhead but pessimistic on recovery latency.

\textbf{(FT6) $k$-replica state redundancy.}
State is written to $k$ peer replicas on each save ($k=2$), allowing
any surviving replica to supply recovery state. FT6 provides state
redundancy rather than Reed--Solomon gradient
coding~\cite{tandon2017gradient,lee2017coded}; our harness therefore
does not perform coded-gradient reconstruction or a head-to-head
gradient-coding comparison.

\subsubsection*{Reconfiguration and baseline (FT7--FT9)}

\textbf{(FT7) Data redistribution.}
After failure, surviving ranks rebalance their data partitions and
continue at a smaller world size. This applies only to architectures
with per-rank data shards.

\textbf{(FT8) Elastic process group.}
PyTorch Elastic~\cite{pytorch_elastic} and Horovod Elastic dynamically
re-form the process group after worker loss, allowing training to
resume on the reduced cluster without application-level intervention.

\textbf{(FT9) No-FT baseline.}
FT9 attaches no fault-tolerance machinery at all: every hook is a
no-op and recovery is declined, so the runner propagates the
underlying training exception. It is the control against which the
other eight mechanisms are measured, and it is what exposes each
architecture's inherent failure behaviour in RQ1. Asynchronous
gradient submission without a barrier~\cite{li2014scaling,recht2011hogwild}
is a property of the A1-async \emph{architecture} rather than an FT
mechanism, and is treated as such throughout.

We call FT1--FT8 mechanisms and FT9 the baseline; the
compatibility matrix records which architecture--mechanism pairs are
viable, degenerate, or structurally infeasible.

\section{Related Work}
\label{sec:related}
We group prior crash-FT work into checkpointing, redundancy
and coding, and decentralized or elastic optimization.

\subsection{Fault Tolerance in Distributed Deep Learning}
\label{sec:related:crash}
Survey treatments catalog the design space without reporting
empirical
comparisons~\cite{cappello2014toward,shi2020fault,khan2024hierarchical};
large-scale production
reports~\cite{zhang2022opt,llama3card,jiang2024megascale,wu2024falcon}
motivate the problem but do not themselves compare mechanisms.

\textbf{Checkpoint-based recovery.}
Checkpoint--restart has been the dominant fault-tolerance strategy in
HPC since the 1970s, and the classical theory of optimal checkpoint
placement under exponential failure arrivals is due to Young and
Daly~\cite{young1974first,daly2006higher}. General-purpose transparent
checkpointing tools such as DMTCP~\cite{ansel2009dmtcp} predate the
deep-learning era and serve as baselines for DDL-specific systems.
Within DDL, modern work targets the multi-terabyte state of large
models. CheckFreq~\cite{mohan2021checkfreq} adapts the checkpoint
interval to keep persistence cost near zero.
Gemini~\cite{wang2023gemini} replicates the checkpoint into peer GPU
and host memory to make restoration near-instantaneous.
DataStates-LLM~\cite{maurya2024datastates} and
ByteCheckpoint~\cite{wan2024bytecheckpoint} push the save path off
the critical path through asynchronous and lazy persistence, and
just-in-time checkpointing~\cite{gupta2024jit} captures state only
when a preemption signal arrives. Each of these systems is benchmarked
against other checkpoint variants on the architecture and workload
that showcase its design choice.

\textbf{Proactive redundancy and coded computing.}
A second group of systems pre-provisions redundancy so that a failed
worker's role is taken over immediately instead of being recovered
from a checkpoint. Bamboo~\cite{thorpe2023bamboo} and
Oobleck~\cite{jang2023oobleck} replicate pipeline stages onto pre-warm
backups. Varuna~\cite{athlur2022varuna} layers redundant scheduling on
top of cloud preemption, and SWIFT~\cite{li2023swift} explores
redundancy in the tensor-parallel setting. At the algorithm layer,
gradient coding~\cite{tandon2017gradient} and coded matrix
multiplication~\cite{lee2017coded} apply erasure codes to worker
gradients so that any sufficient subset reconstructs the aggregate
update, providing straggler tolerance with provable bounds. Each of
these systems is evaluated against vanilla parallelism in the regime
the system targets; cross-comparison against checkpoint-class
mechanisms is uncommon.

\textbf{Benchmarks and fault injection.}
Community benchmarks such as MLPerf~\cite{mattson2020mlperf}
standardize \emph{performance} comparison across training
systems but do not inject faults or characterize recovery;
general-purpose chaos-engineering tools such as Chaos
Mesh~\cite{chaosmesh} inject faults but are agnostic to
training semantics and report no training-specific metrics.
\sysname{} occupies the intersection: MLPerf-style fixed
configuration disclosure combined with controlled fault
schedules and recovery-specific metrics. Orthogonally, a
line of work addresses \emph{non-crash} failure models:
Byzantine-robust
aggregation~\cite{blanchard2017machine,yin2018byzantine,karimireddy2021byzantine,el2018hidden}
and silent data
corruption~\cite{dixit2021silent,hochschild2021cores}.
These we deliberately scope out: \sysname{} injects fail-stop
crashes only, and its mechanisms assume sane survivors.

\textbf{Decentralized, asynchronous, and elastic optimization.}
A third group avoids tightly-coupled synchronization in the first
place, accepting some convergence cost in exchange for graceful
degradation. Decentralized parallel SGD~\cite{lian2017can} replaces
the all-reduce with peer-to-peer averaging over a gossip overlay, and
subsequent theoretical work bounds the convergence penalty under
topology changes and worker
failures~\cite{koloskova2020unified,assran2019stochastic}. Asynchronous
parameter servers~\cite{li2014scaling} and
Hogwild!~\cite{recht2011hogwild} tolerate stragglers and crashes by
design at the cost of stale gradients. PyTorch
Elastic~\cite{pytorch_elastic} and Horovod
Elastic~\cite{sergeev2018horovod} expose explicit
process-group reconfiguration APIs so the cluster can grow and shrink
during training; production frameworks such as
Megatron-LM~\cite{shoeybi2019megatron} and
DeepSpeed~\cite{rasley2020deepspeed} layer elasticity hooks on top of
these primitives.

The works above typically evaluate mechanisms within the design space
they target: checkpoint work against other checkpoint variants,
redundancy work against vanilla parallelism, and gossip-SGD within the
peer-to-peer overlay. In this paper, we instead vary both the training
architecture and fault-tolerance mechanism on the same cluster (the
framework section describes the 7$\times$9 compatibility matrix and
142 reported tuples). Gradient coding requires coded-gradient arithmetic
and a compatible aggregation topology that our harness does not
implement, so we do not report a head-to-head comparison. The
7$\times$9 matrix further distinguishes viable, degenerate, and
structurally infeasible (S) architecture--mechanism pairs. The latter
are explicitly considered but cannot execute without redesigning the
architecture or mechanism, exposing compatibility constraints that
per-mechanism evaluations do not systematically characterize.
\section{The \sysname{} Framework}
\label{sec:framework}

\sysname{} pairs distributed-training architectures with
fault-tolerance mechanisms under a common fault injector and
metric collector. The implementation is released as an artifact.

\subsection{Design Goals}

Four goals guide the design: \emph{common interfaces}, with
architectures and mechanisms implemented as plug-ins behind two
abstract base classes; \emph{determinism}, where a seed and fault
schedule reproduce a run up to CUDA floating-point nondeterminism;
\emph{architecture-agnostic injection}, where faults target ranks
rather than framework internals; and \emph{comparable metrics},
collected under consistent definitions.

\subsection{Architecture--Mechanism Compatibility Matrix}
\label{sec:framework:matrix}

\begin{table}[h]
\caption{Architecture--mechanism compatibility matrix:
\texttt{V} viable, \texttt{D} degenerate, and \texttt{S}
structurally infeasible.}
\label{tab:matrix}
\centering
\small
\setlength{\tabcolsep}{5pt}
\begin{adjustbox}{max width=9cm}
\begin{tabular}{@{}lccccccccc@{}}
\toprule
 & FT1 & FT2 & FT3 & FT4 & FT5 & FT6 & FT7 & FT8 & FT9 \\
 & Per & Mem & Async & JIT & Repl & $k$-repl & Redist & Elastic & No-FT \\
\midrule
A1 PS        & V & V & V & V & V & V & V & V & V \\
A2 Ring-AR   & V & V & V & V & V & V & V & V & V \\
A3 TreeAR    & V & V & V & V & V & V & V & V & V \\
A4 Pipeline  & V & V & V & V & V & V & D & V & V \\
A5 Tensor    & V & V & V & V & V & V & D & V & V \\
A6 P2P       & V & S & V & V & S & S & V & V & V \\
A7 3D Hybrid & V & V & V & V & V & V & D & V & V \\
\bottomrule
\end{tabular}
\end{adjustbox}
\end{table}

Table~\ref{tab:matrix} defines the architecture--mechanism
configuration space. Each pair is classified as:

\begin{itemize}
\item \textbf{\texttt{V} -- viable.} The mechanism can be implemented
as designed and evaluated by \sysname{}.

\item \textbf{\texttt{D} -- degenerate.} The pair is implementable,
but the mechanism collapses into a no-op or an already measured
alternative. For example, FT7 on A4, A5, and A7 has no per-rank
data shard to redistribute and reduces to checkpoint-based restart.

\item \textbf{\texttt{S} -- structurally infeasible.} The mechanism's
communication protocol is incompatible with the architecture's
process-group topology without redesign. For example, FT2 requires
a globally addressable mirror rank, whereas A6 exposes only per-edge
gossip subgroups.
\end{itemize}

Across the three fault traces, the matrix yields 180 candidate
(architecture, mechanism, trace) tuples after excluding degenerate
cells; \sysname{} reports 142. Coverage and measurement provenance
(\markm{} versus \marka{}) are detailed in the methodology.

The matrix therefore exposes whether a mechanism remains viable,
becomes degenerate, or is structurally infeasible when transferred
across training architectures.

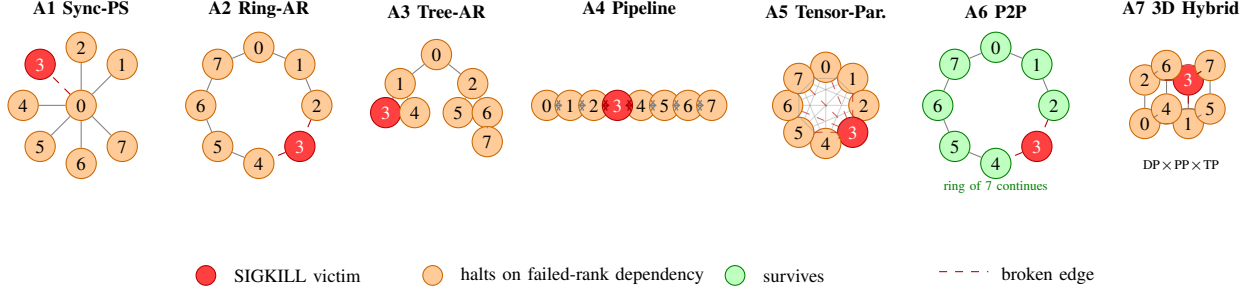
\begin{figure*}[t]
\centering
\begin{tikzpicture}[
  rk/.style   ={circle, draw=black, fill=white, minimum size=4mm,
                inner sep=0pt, font=\scriptsize},
  dead/.style ={circle, draw=red!70!black, fill=red!75, minimum size=4mm,
                inner sep=0pt, font=\scriptsize, text=white},
  halt/.style ={circle, draw=orange!80!black, fill=orange!40, minimum size=4mm,
                inner sep=0pt, font=\scriptsize},
  ok/.style   ={circle, draw=green!50!black, fill=green!25, minimum size=4mm,
                inner sep=0pt, font=\scriptsize},
  edge/.style ={-, thin, gray},
  cut/.style  ={dashed, red!70!black, thin},
  panel title/.style={font=\scriptsize\bfseries, anchor=south},
]


\begin{scope}[xshift=0cm,yshift=0cm,scale=0.48]
  \node[panel title] at (0,2.15) {A1 Sync-PS};

  \node[halt] (a10) at (0,0) {0};

  \foreach \k/\ang in {
    1/45,
    2/90,
    3/135,
    4/180,
    5/225,
    6/270,
    7/315}{
    \ifnum\k=3
      \node[dead] (a1\k) at (\ang:1.6) {\k};
    \else
      \node[halt] (a1\k) at (\ang:1.6) {\k};
    \fi
  }

  \foreach \k in {1,2,4,5,6,7}{
    \draw[edge] (a10) -- (a1\k);
  }

  \draw[cut] (a10) -- (a13);
\end{scope}

\begin{scope}[xshift=2.35cm,yshift=0cm,scale=0.48]
  \node[panel title] at (0,2.15) {A2 Ring-AR};

  \foreach \k/\ang in {
    0/90,
    1/45,
    2/0,
    3/-45,
    4/-90,
    5/-135,
    6/180,
    7/135}{
    \ifnum\k=3
      \node[dead] (a2\k) at (\ang:1.6) {\k};
    \else
      \node[halt] (a2\k) at (\ang:1.6) {\k};
    \fi
  }

  \foreach \i/\j in {
    0/1,
    1/2,
    2/3,
    3/4,
    4/5,
    5/6,
    6/7,
    7/0}{
    \ifnum\i=2
      \draw[cut] (a2\i)--(a2\j);
    \else
      \ifnum\i=3
        \draw[cut] (a2\i)--(a2\j);
      \else
        \draw[edge] (a2\i)--(a2\j);
      \fi
    \fi
  }
\end{scope}

\begin{scope}[xshift=4.70cm,yshift=0cm,scale=0.48]
  \node[panel title] at (0,2.15) {A3 Tree-AR};

  \node[halt] (a30) at ( 0,   1.4) {0};
  \node[halt] (a31) at (-1,   0.6) {1};
  \node[halt] (a32) at ( 1,   0.6) {2};

  \node[dead] (a33) at (-1.4,-0.2) {3};
  \node[halt] (a34) at (-0.6,-0.2) {4};
  \node[halt] (a35) at ( 0.6,-0.2) {5};
  \node[halt] (a36) at ( 1.4,-0.2) {6};
  \node[halt] (a37) at ( 1.4,-1.0) {7};

  \draw[edge] (a30) -- (a31);
  \draw[edge] (a30) -- (a32);

  \draw[cut]  (a31) -- (a33);
  \draw[edge] (a31) -- (a34);

  \draw[edge] (a32) -- (a35);
  \draw[edge] (a32) -- (a36);

  \draw[edge] (a36) -- (a37);
\end{scope}

\begin{scope}[xshift=7.25cm,yshift=0cm,scale=0.48]
  \node[panel title] at (0,2.15) {A4 Pipeline};

  \foreach \k in {0,1,2,4,5,6,7}{
    \node[halt] (a4\k) at ({\k*0.65-2.275},0) {\k};
  }

  \node[dead] (a43) at (-0.325,0) {3};

  \foreach \i/\j in {
    0/1,
    1/2,
    2/3,
    3/4,
    4/5,
    5/6,
    6/7}{
    \ifnum\i=2
      \draw[<->,dashed,red!70!black,semithick]
        (a4\i)--(a4\j);
    \else
      \ifnum\i=3
        \draw[<->,dashed,red!70!black,semithick]
          (a4\i)--(a4\j);
      \else
        \draw[<->,semithick,gray]
          (a4\i)--(a4\j);
      \fi
    \fi
  }
\end{scope}

\begin{scope}[xshift=9.85cm,yshift=0cm,scale=0.48]
  \node[panel title] at (0,2.15) {A5 Tensor-Par.};

  \foreach \k/\ang in {
    0/90,
    1/45,
    2/0,
    3/-45,
    4/-90,
    5/-135,
    6/180,
    7/135}{
    \ifnum\k=3
      \node[dead] (a5\k) at (\ang:1.05) {\k};
    \else
      \node[halt] (a5\k) at (\ang:1.05) {\k};
    \fi
  }

  \foreach \i in {0,1,2,4,5,6,7}{
    \foreach \j in {0,1,2,4,5,6,7}{
      \ifnum\i<\j
        \draw[edge,opacity=0.35]
          (a5\i)--(a5\j);
      \fi
    }
  }

  \foreach \k in {0,1,2,4,5,6,7}{
    \draw[cut,opacity=0.6]
      (a5\k)--(a53);
  }
\end{scope}

\begin{scope}[xshift=12.10cm,yshift=0cm,scale=0.48]
  \node[panel title] at (0,2.15) {A6 P2P};

  \foreach \k/\ang in {
    0/90,
    1/45,
    2/0,
    3/-45,
    4/-90,
    5/-135,
    6/180,
    7/135}{
    \ifnum\k=3
      \node[dead] (a6\k) at (\ang:1.6) {\k};
    \else
      \node[ok] (a6\k) at (\ang:1.6) {\k};
    \fi
  }

  \foreach \i/\j in {
    0/1,
    1/2,
    2/3,
    3/4,
    4/5,
    5/6,
    6/7,
    7/0}{
    \ifnum\i=2
      \draw[cut] (a6\i)--(a6\j);
    \else
      \ifnum\i=3
        \draw[cut] (a6\i)--(a6\j);
      \else
        \draw[edge] (a6\i)--(a6\j);
      \fi
    \fi
  }

  \node[
    font=\tiny,
    green!50!black,
    align=center
  ] at (0,-2.25)
  {ring of 7 continues};
\end{scope}

\begin{scope}[xshift=14.55cm,yshift=0cm,scale=0.48]
  \node[panel title] at (0,2.15) {A7 3D Hybrid};

  \node[halt] (a70) at (-1,-0.5) {0};
  \node[halt] (a71) at ( 0.2,-0.5) {1};
  \node[halt] (a72) at (-1, 0.7) {2};
  \node[dead] (a73) at ( 0.2,0.7) {3};

  \node[halt] (a74) at (-0.4,-0.1) {4};
  \node[halt] (a75) at ( 0.8,-0.1) {5};
  \node[halt] (a76) at (-0.4, 1.1) {6};
  \node[halt] (a77) at ( 0.8, 1.1) {7};

  \draw[edge] (a70)--(a71);
  \draw[edge] (a72)--(a73);
  \draw[edge] (a74)--(a75);
  \draw[edge] (a76)--(a77);

  \draw[edge] (a70)--(a72);
  \draw[edge] (a71)--(a73);
  \draw[edge] (a74)--(a76);
  \draw[edge] (a75)--(a77);

  \draw[edge] (a70)--(a74);
  \draw[edge] (a71)--(a75);
  \draw[edge] (a72)--(a76);
  \draw[edge] (a73)--(a77);

  \draw[cut] (a73)--(a72);
  \draw[cut] (a73)--(a71);
  \draw[cut] (a73)--(a77);

  \node[font=\tiny] at (0,-1.65)
  {DP$\times$PP$\times$TP};
\end{scope}

\begin{scope}[xshift=1.65cm,yshift=-2.25cm]

  \node[dead,scale=0.65] (l1) at (0,0) {};
  \node[anchor=west,font=\scriptsize]
    at (0.25,0)
    {SIGKILL victim};

  \node[halt,scale=0.65] (l2) at (3.0,0) {};
  \node[anchor=west,font=\scriptsize]
    at (3.25,0)
    {halts on failed-rank dependency};

  \node[ok,scale=0.65] (l3) at (7.0,0) {};
  \node[anchor=west,font=\scriptsize]
    at (7.25,0)
    {survives};

  \draw[cut] (9.7,0.05) -- (10.35,0.05);
  \node[anchor=west,font=\scriptsize]
    at (10.4,0)
    {broken edge};

\end{scope}

\end{tikzpicture}

\caption{
Per-architecture failure topology when rank 3 is SIGKILL'd.
Orange ranks halt when they next require the failed rank, green ranks
continue through graceful degradation, and dashed red lines mark
broken edges.
}
\label{fig:failure-topology}
\vspace{-0.3cm}

\end{figure*}

\subsection{Fault Injector}
\label{sec:framework:injector}

The injector executes deterministic, architecture-agnostic fault
schedules by issuing an in-process SIGKILL to the target rank.
The schedule format, crash traces, and fault-model scope are defined
in the methodology. Figure~\ref{fig:failure-topology} illustrates
the resulting failure topology under F1.

Per-step measurements and fault events are logged in
\texttt{jsonl} and aggregated offline; metric definitions are
given in the methodology.



\section{Experimental Methodology}
\label{sec:methodology}
This section describes the execution platform, fault-injection
protocol, workloads, sweep parameters, metrics, and statistical
procedures used in the evaluation.

\subsection{Execution Platform: Matilda HPC}
\label{sec:methods:cluster}
All experiments run on Oakland University's Matilda HPC cluster.
Each compute node has four NVIDIA Tesla V100 PCIe 16\,GB GPUs,
32 CPU cores, and 192\,GB of host memory, interconnected by HDR
InfiniBand and BeeGFS. NCCL ring all-reduce achieves 9.7\,GB/s
effective bus bandwidth on 8 GPUs across 2 nodes (16\,MiB tensors,
100 iterations). Jobs are submitted through SLURM, with one array
task per (architecture, mechanism, fault-trace, seed) cell.

The software stack comprises PyTorch 2.3.0, CUDA 12.1,
NCCL 2.20.5, and Python 3.10. All architectures run as
\sysname{} plug-ins over this stack.

\subsection{Fault Injection Techniques}
\label{sec:methods:faults}
The \sysname{} injector executes a JSON-described fault schedule
of \texttt{(time, rank, type, params)} tuples. We use three fixed
crash traces:
\begin{itemize}
\item \textbf{F1 -- Single fail-stop crash.}
One worker rank receives
\texttt{os.kill(getpid(), SIGKILL)} from its injector daemon at
$t\!\in\![30,120]$\,s, and at $t\!=\!180$\,s for A6. Injection is
delayed for large-state architectures so JIT mechanisms can complete
their emergency save; survivors observe the failure through their
architecture-specific communication or detection path.

\item \textbf{F2 -- Concurrent crashes.}
Two simultaneous F1 events target ranks $r$ and $r{+}2$,
exposing single-replica ring mechanisms to replica collision.

\item \textbf{F3 -- Cascading crashes.}
Three F1 events occur at $t$, $t{+}30$\,s, and $t{+}60$\,s;
TTD and recovery latency are recorded separately for each fault.
\end{itemize}

The injector runs as a daemon thread in each rank's runner and
dispatches events in registration order: the metric collector logs
the event, the recovery supervisor marks it pending, the architecture
receives \texttt{on\_failure} so eager-membership schemes can drop the
failed rank, FT mechanism prepares for recovery, and the executor
applies the fault on the target rank.

The injector is architecture-agnostic: the same OS-level action is
used across frameworks, with fixed fault type, rank selection, fault
count, and inter-arrival spacing across all seven architectures.
Only injection time varies with per-rank state size; A6 uses
$t\!=\!180$\,s because its per-edge subgroups require $\sim$60\,s
to stabilize. We assume the SLURM scheduler, NCCL, and kernel TCP/IB
stack are correct.

\subsection{Workloads}
\label{sec:methods:workloads}

Workloads are chosen per architecture family to preserve relevant
inter rank tensor shapes. Data parallel architectures
(A1, A2, A3, A6) train \textbf{ResNet 50} on synthetic data with
micro batch 32 per rank and fp32 SGD with momentum 0.9, yielding
approximately 200\,MB of state per rank. To exercise A4, A5, and A7
under realistic pipeline, tensor, and 3D parallel settings, we use
Megatron style MLP stacks with 2.5 to 2.7\,GB of state per rank.

The MLPs reproduce the Transformer feed forward sublayer through
column parallel expansion, element wise nonlinearity, and row parallel
contraction. They preserve properties relevant to fault tolerance,
including per rank state size, tensor shapes, communication volume,
collectives, and process group topology, while omitting attention,
embeddings, and sequence processing. Thus, they capture the
communication and state behavior of large Transformer training
without implementing a complete language model.

\subsection{Sweep Parameters}
\label{sec:methods:sweep}
All experiments use 8 V100 GPUs across two nodes. Across the study,
viable configurations are evaluated under crash traces F1, F2, and F3,
with architecture specific layouts but fixed GPU count: 1 PS and
7 workers for A1, an 8 rank ring for A2, an 8-rank tree for A3,
an 8 stage pipeline for A4, a TP group of 8 for A5, an 8 peer
gossip overlay for A6, and a
DP$\times$PP$\times$TP$\!=\!2\!\times\!2\!\times\!2$ mesh for A7.

\subsection{Statistical Protocol}
\label{sec:methods:stats}

Sample sizes range from $n\!=\!1$ to $7$, depending on wall clock
cost and experiment type. Structurally infeasible (\texttt{S}) cells
are confirmed by smoke runs and provide no viable
\ensuremath{T_{\text{restore}}}.

FT7 and FT8 activate only after a failure and thus add no steady
state overhead. Their \ensuremath{T_{\text{restore}}} values are
analytical lower bounds rather than direct measurements because
recovery depends on external rendezvous and process reinitialization
that the runner cannot execute, such as PyTorch Elastic or a c10d
rendezvous server.

Crash experiments with no fault tolerance mechanism enabled report
the mean with a 95\% Student's $t$ confidence interval. Fault
tolerance mechanism experiments report the mean and half range,
$(\max-\min)/2$, because their smaller sample sizes do not support
a reliable normality assumption. Outliers are retained, and runs
using the same seed replay the same deterministic fault schedule.

\subsection{Metric Definitions}
\label{sec:methods:metrics}

We report four metrics: steady state overhead (\emph{SS overhead}),
time to detect (\emph{TTD}), effective state restore cost
$T_{\text{restore}}$, and recovery time $T_{\text{recover}}$:
\begin{align}
\textit{SS overhead} &= T_{\text{off}}/T_{\text{on}} - 1,\\
\textit{TTD} &= t_{\text{detected}} - t_{\text{fault}},\\
T_{\text{restore}} &= \text{effective state restore cost},\\
T_{\text{recover}} &= T_{\text{restore}} + T_{\text{PG}} + T_{\text{init}}.
\end{align}

$T_{\text{on}}$ and $T_{\text{off}}$ denote throughput with and
without FT, respectively. Detection is the first failure indication
observed by a surviving rank. $T_{\text{restore}}$ captures the
mechanism's effective restore cost. $T_{\text{PG}}$ is the time to
rebuild the process group after a failure, including rendezvous and
communicator setup, while $T_{\text{init}}$ is the time to
reinitialize the FT mechanism and its required state. Total fault
to resume latency is $\textit{TTD}+T_{\text{recover}}$. Recovery
latency entries are marked \emph{m} when measured and \emph{a}
when analytical.

\subsection{Restore Time Model and Validation}
\label{sec:methods:restore-model}

On the halting architectures, we cannot observe
$T_{\text{restore}}$ after a worker crash. Surviving ranks block
on communication with the failed rank until PyTorch's 600\,s
process-group watchdog fires. The watchdog invokes
\texttt{std::terminate} from a C++ thread, killing the process
before the Python recovery path can log the restore event.
We therefore estimate $T_{\text{restore}}$ from the mechanism's
dominant state transfer path.

For in memory FT2 and the restore component of FT4, the estimate
is state size divided by the PCIe Gen 3 bandwidth
($\sim$12\,GB/s). For disk based FT1 and FT3, it is state size
divided by the measured filesystem read bandwidth. For FT4, the
measured emergency save cost is added separately when reporting
its effective restore cost.
When the recovery supervisor observes training resume,
$T_{\text{recover}}$ is measured directly. Otherwise, recovery
is reported as a lower bound by combining analytical
$T_{\text{restore}}$ with the estimated process group reformation
cost $T_{\text{PG}}$ and mechanism reinitialization cost
$T_{\text{init}}$.

\paragraph{M1 validation of the analytical model}
We validate the analytical restore model with a
microbenchmark covering transfer paths and two state sizes:
approximately 205\,MB for ResNet 50 and 2.6\,GB for the MLP
workload. Each condition uses $n\!=\!10$ cold cache iterations.
Disk measurements load serialized \texttt{.pt} state dictionaries
and include \texttt{torch.load} deserialization, while
PCIe measurements time pinned memory to GPU transfer from a
preloaded buffer.

\begin{table}[h]
\caption{M1 validation of the analytical $T_{\text{restore}}$
model using standalone restore path microbenchmarks
($n\!=\!10$). Deviation is
$(\text{measured}-\text{predicted})/\text{predicted}$.}
\label{tab:m1-validation}
\centering \small
\begin{adjustbox}{max width=\columnwidth}
\begin{tabular}{@{}l r r r r@{}}
\toprule
Condition & State (MB) & Predicted (ms) &
Measured (ms $\pm$ CI95) & Deviation \\
\midrule
PCIe / small state & 204.7  & $17.1$   & $20.5 \pm 0.7$    & $+20.2\%$ \\
PCIe / large state & 2600.0 & $216.7$  & $289.8 \pm 0.9$   & $+33.7\%$ \\
Disk / large state & 2600.0 & $1968.2$ & $3270.5 \pm 70.3$ & $+66.2\%$ \\
Disk / small state & 204.7  & $155.0$  & $340.2 \pm 5.2$   & $+119.5\%$ \\
\bottomrule
\end{tabular}
\end{adjustbox}
\end{table}

M1 validates the analytical model as a conservative lower bound
and quantifies the gap between bandwidth based estimates and
realized restore time. Measured restore time is approximately
20\%--120\% above the analytical bound. The smaller PCIe gap
reflects transfer and synchronization overhead, whereas the larger
disk gap also includes \texttt{torch.load} deserialization costs,
which are particularly significant for small checkpoints containing
many tensors. We therefore treat analytical $T_{\text{restore}}$
values throughout the paper as lower bounds rather than exact
restore times.

\subsection{Measurement Provenance Across the 142 Tuples}
\label{sec:methods:provenance}

The 142 reported tuples comprise 133 viable and 9 structurally
infeasible (\texttt{S}) cases. Table~\ref{tab:provenance}
summarizes their measurement provenance. Of the viable tuples,
10 have directly observed post fault recovery or continuation,
while 123 use analytical $T_{\text{restore}}$ or reconfiguration
bounds. Steady state overhead is measured where applicable and
is zero by construction for reactive only mechanisms such as
FT7 and FT8. All 9 \texttt{S} cases halted in smoke tests due to structural infeasibility.

\begin{table}[h]
\caption{Measurement provenance of the 142 reported tuples.}
\label{tab:provenance}
\centering
\small
\begin{tabular}{@{}l r l@{}}
\toprule
Provenance & $N$ & Recovery evidence \\
\midrule
Direct observation      & 10  & Measured \\
Analytical bound         & 123 & Analytical \\
Structurally infeasible  & 9   & No viable recovery \\
\midrule
\textbf{Total}           & \textbf{142} & \\
\bottomrule
\end{tabular}
\vspace{-0.3cm}
\end{table}

The 10 directly observed tuples are A6+FT1/FT4 under F1--F3,
A6+FT9 F1, A1 async+FT9 F1, A5+FT4 F2, and A7+FT4 F2.
The reported set covers 133 viable tuples across F1--F3, split
as 43 (F1), 46 (F2) and 53 (F3) unique (architecture,
mechanism) pairs at $n\!\in\!\{1,2\}$ seeds per cell and
boosted to $n\!=\!5$ on the seven cells cited by the decision
framework, together with all 9 structurally infeasible cases.
The F1 count is the smallest because a subset of FT7/FT8 seed
sets failed during warmup and never reached steady state.
\section{RQ1: Inherent Failure Exposure in Distributed Deep Learning Architectures}
\label{sec:rq1}

\subsection{Motivation}
\label{sec:rq1:motivation}

Before evaluating fault-tolerance mechanisms, we first
characterize each architecture's inherent response to worker
loss: whether surviving ranks can continue training and how
quickly they observe the failure. This no-FT baseline establishes
what a recovery mechanism must improve upon and reveals how
the communication topology itself shapes failure exposure.

\subsection{Approach}
\label{sec:rq1:approach}

For each architecture, we run the corresponding workload
without an FT mechanism and inject F1, a single SIGKILL of
one worker rank. The fault is injected at $t\!=\!30$\,s, except
for A6, where it is delayed to $t\!=\!180$\,s so its per-edge
subgroups reach stable pre-fault throughput. We record whether
surviving ranks continue training, time-to-detect (TTD),
post-fault throughput, and final job outcome. TTD is the interval
from SIGKILL to the first failure indication observed by a
surviving rank or failure detector. Sample sizes are $n\!=\!5$
for the six halting configurations, $n\!=\!3$ for A1-async,
and $n\!=\!7$ for A6.

Table~\ref{tab:rq1-results} summarizes the F1 measurements:
pre-fault throughput, TTD, post-fault throughput, aggregate
slowdown, and whether training halts or degrades gracefully.

\subsection{Results}
\label{sec:rq1:results}

\begin{table*}[t]
\caption{F1 exposure on 8$\times$V100
(SIGKILL at $t\!=\!30$\,s; A6 at $t\!=\!180$\,s).
Workloads are ResNet-50 for A1--A3/A6 and Megatron style
MLPs for A4/A5/A7; throughput is comparable only within
workload family. Values are mean $\pm$95\% CI.
Aggregate slowdown is
$1-T_{\text{post}}N_{\text{post}}/
(T_{\text{pre}}N_{\text{pre}})$, with $N$ active workers;
negative values indicate higher post fault aggregate throughput.}
\label{tab:rq1-results}
\centering
\small
\setlength{\tabcolsep}{4pt}
\begin{adjustbox}{max width=\textwidth}
\begin{tabular}{@{}l c l c c c c c@{}}
\toprule
ID & $n$ & Backend & Pre-fault thpt & TTD & Halts? &
Post-fault thpt & Agg.\ slowdown \\
   &     &         & (samples/s/rank) & (s) & &
(samples/s/rank) & (\%) \\
\midrule
A1 Synchronous PS
& 5 & NCCL+IB
& $174.9 \pm 0.4$
& $600.6 \pm 0.3$
& yes & 0 (halted) & 100 \\

A1 Asynchronous PS
& 3 & gloo+TCP
& $32.4 \pm 0.3$
& $5.26 \pm 0.01$
& no & $32.6 \pm 0.6$
& $13.65 \pm 1.34$ \\

A2 Ring All-Reduce
& 5 & NCCL+IB (ring)
& $284.9 \pm 0.1$
& $601.7 \pm 1.8$
& yes & 0 (halted) & 100 \\

A3 Tree All-Reduce
& 5 & NCCL+IB (tree)
& $277.1 \pm 0.1$
& $605.9 \pm 0.2$
& yes & 0 (halted) & 100 \\

A4 Pipeline Parallelism
& 5 & NCCL+IB (send/recv)
& $951.5 \pm 3.2$
& $600.13 \pm 0.03$
& yes & 0 (halted) & 100 \\

A5 Tensor Parallelism
& 5 & NCCL+IB (all-reduce)
& $1013.5 \pm 0.6$
& $604.21 \pm 1.50$
& yes & 0 (halted) & 100 \\

A6 P2P / D-PSGD (gossip)
& 7 & gloo+TCP
& $10.68 \pm 0.59$
& $\approx 0.5$
& no
& $14.38 \pm 5.53$
& $-17.87$ (aggregate up) \\

A7 3D Hybrid (DP$\times$PP$\times$TP)
& 5 & NCCL+IB (3 axes)
& $85.9 \pm 0.1$
& $615.24 \pm 0.58$
& yes & 0 (halted) & 100 \\
\bottomrule
\end{tabular}
\end{adjustbox}
\vspace{-0.3cm}
\end{table*}

\textbf{Configurations that depend on the failed rank halt
after a single worker failure.}
A1-sync, A2, A3, A4, A5, and A7 halt under F1, with no
surviving seed resuming training. In each case, surviving ranks
eventually block on communication that requires the failed rank:
parameter-server synchronization in A1-sync, all-reduce in
A2/A3/A5, pipeline communication in A4, and one or more of
the DP, PP, and TP communication paths in A7. Their TTDs
cluster around PyTorch's default 600\,s process-group watchdog,
ranging from $600.13\pm0.03$\,s on A4 to
$615.24\pm0.58$\,s on A7. Thus, the observed detection delay
is primarily a runtime timeout effect rather than an inherent
property of the training topology.

\textbf{A1-async continues under F1 with a permissive
responder threshold.}
Unlike A1-sync, the asynchronous parameter server uses
non-blocking worker receives under a 5\,s timeout and
averages whatever gradients arrive within that window. We set the
responder threshold to one, so a round completes as long as any
single worker replies and the server never waits on a quorum.
Under F1, the server detects the missing worker in
$5.26\pm0.01$\,s, removes it from the live set, and continues
with the remaining six workers. All three seeds complete the
1500-step run. Aggregate throughput decreases from
$226.8\pm2.1$ to $195.9\pm3.4$ samples/s, corresponding to a
$13.65\%\pm1.34\%$ slowdown, close to the structural loss
$1-6/7\approx14.3\%$. The shorter TTD therefore comes from
the asynchronous receive timeout rather than from the
global process-group watchdog.

\textbf{The 600\,s TTD is controlled by the process-group
timeout.}
To isolate the source of the long detection latency, we rerun
A2 under F1 with process-group timeouts of 30, 120, and
600\,s. As Table~\ref{tab:rq1-watchdog-sensitivity} shows,
measured TTD follows the configured timeout almost exactly:
$30.01\pm0.01$, $120.03\pm0.01$, and
$600.03\pm0.01$\,s, respectively. The approximately 600\,s
TTD observed across the six halting configurations is therefore
a consequence of PyTorch's default process-group timeout, not
an architecture-imposed detection floor. Deployments that pass
a shorter \texttt{timeout} to
\texttt{torch.distributed.init\_process\_group} obtain
proportionally faster failure detection. The
\texttt{NCCL\_TIMEOUT} environment variable does \emph{not}
control this watchdog: an earlier sweep over
$\{30, 120, 600\}$\,s that left the
\texttt{init\_process\_group} argument unchanged still produced
$\sim$600\,s TTD on every seed, because the watchdog timeout is
fixed when the process group is constructed.

\begin{table}[h]
\caption{A2 Ring-AR + F1 no-FT baseline at three
\texttt{pg\_timeout} settings, showing that TTD tracks the
configured timeout.}
\label{tab:rq1-watchdog-sensitivity}
\centering
\small
\begin{tabular}{@{}r r r r@{}}
\toprule
\texttt{pg\_timeout} & $n$ & Measured TTD (s) & TTD / setting \\
\midrule
30\,s  & 3 & $30.01\pm0.01$  & $1.00\times$ \\
120\,s & 3 & $120.03\pm0.01$ & $1.00\times$ \\
600\,s & 3 & $600.03\pm0.01$ & $1.00\times$ \\
\bottomrule
\end{tabular}
\vspace{-0.1cm}

\end{table}

\textbf{A6 degrades gracefully, but higher sample throughput
does not improve loss progress.}
A6 has no global collective: each rank communicates only with
its live gossip neighbors and receives failure information
through membership notification. Because its per-edge
subgroups require up to $\sim$60\,s to reach a stable timing
regime, F1 is injected at $t\!=\!180$\,s. The failure is observed
within approximately 500\,ms, after which the two neighbors of
the failed rank continue with one gossip partner while the
remaining ranks retain two.

Mean per-rank throughput rises from
$10.68\pm0.59$ to $14.38\pm5.53$ samples/s after the failure.
Consequently, aggregate sample throughput increases by
17.9\%, from 85.4 to 100.7 samples/s. The gain is
concentrated where the topology predicts it: the two ranks
adjacent to the failed worker now wait on a single gossip
partner per round and run at $2\times$ their pre-fault
rate, while the five non-adjacent ranks stay near baseline.
This distribution is consistent with reduced synchronization
pressure; given the wide confidence interval at $n\!=\!7$, we
treat the 17.9\% aggregate increase as descriptive.

\begin{figure}[h]
\centering
\includegraphics[width=0.95\columnwidth]{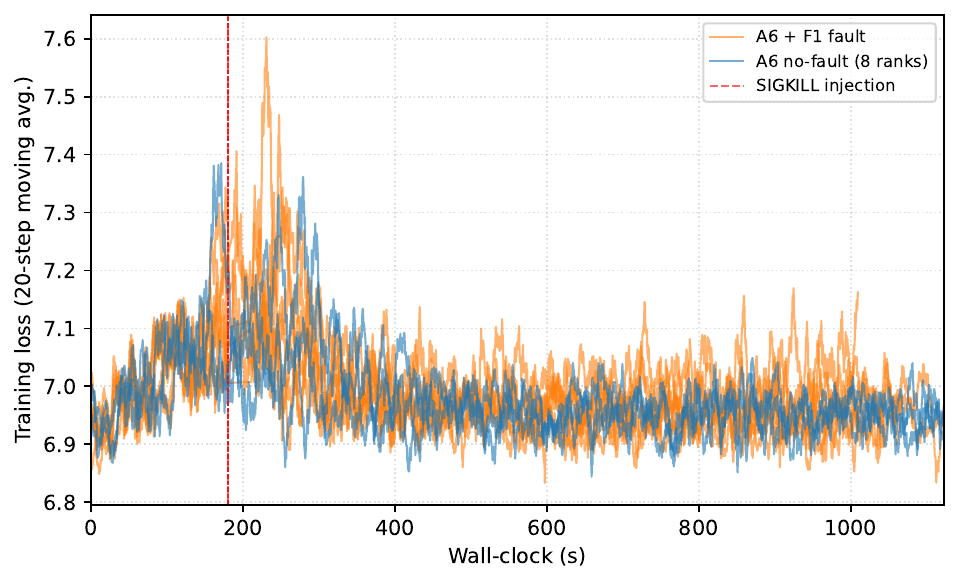}
\vspace{-0.3cm}
\caption{A6 P2P gossip training loss (20-step moving average)
for seven F1 seeds and three eight-rank no-fault seeds over a
common 1123\,s window. The dashed line marks the SIGKILL at
$t\!=\!180$\,s.}
\label{fig:a6-convergence}
\end{figure}

The throughput increase does not translate into faster
optimization. Measured from the fault over a window matched
against the eight-rank no-fault baseline, the fault runs reduce
loss by $0.130$ on average and the baseline by $0.126$, with no
statistically detectable difference despite the fault runs
processing roughly $18\%$ more samples per second.
Figure~\ref{fig:a6-convergence} shows a transient loss increase
shortly after the fault, but the no-fault baseline produces one
of the same size ($0.30$ against $0.32$\,nats, both peaking
near $t\!=\!220$\,s), so it reflects gossip overlay dynamics on
this workload rather than recovery from the fault. A6
demonstrates graceful execution after worker loss,
but its post-fault throughput gain reflects reduced
communication pressure rather than improved learning
efficiency.

\section{RQ2: Resilience--Cost Trade-offs}
\label{sec:rq2}

\subsection{Motivation}
\label{sec:rq2:motivation}
Once an architecture is fixed, choosing an FT mechanism is a
two-axis trade-off: continuous \emph{steady-state overhead}
(storage, network, redundant compute) during normal training
versus \emph{recovery latency} (\ensuremath{T_{\text{restore}}}) once a fault fires.
Without a quantitative view of how alternative mechanisms
compare on these two axes, the practitioner cannot tell whether
two mechanisms are substitutes or complements, or whether a
given mechanism is even Pareto-competitive. Existing FT papers
each report a favorable single point on this trade-off for the
mechanism they
propose~\cite{wang2023gemini,thorpe2023bamboo,tandon2017gradient,gupta2024jit};
head-to-head comparisons across mechanism families on the same
workload are uncommon. RQ2 reconstructs this frontier
empirically on a fixed deployment.

\subsection{Approach}
\label{sec:rq2:approach}
We hold every choice except the FT mechanism constant.
Workload: ResNet-50 on a synthetic dataset, micro-batch 32,
fp32 SGD with momentum ($\sim$205\,MB per-rank state).
Cluster: 8$\times$V100 across 2 nodes. Architecture: A2
Ring-AR; disk-based mechanisms use BeeGFS scratch. Fault
trace: F1, a single SIGKILL fail-stop at $t\!=\!30$\,s.
On this deployment we evaluate every configuration marked
\texttt{V} in Table~\ref{tab:matrix}.

Per-cell seed count $n$ is scaled to wall-clock cost:
$n\!=\!6$ for FT2, FT3, FT5 and FT6, where seed variance is
informative, and $n\!=\!3$ for FT1 and FT4. FT7 (data redistribution) and
FT8 (elastic reconfiguration) are reactive-only and reported
analytically. We log SS
overhead (relative throughput loss vs.\ the failure-free
RQ1 baseline, Table~\ref{tab:rq1-results}), effective \ensuremath{T_{\text{restore}}},
per-save cost, and expected lost work, then place each configuration on the
(SS, \ensuremath{T_{\text{restore}}}) plane and identify the frontier.

We run two nested experiments. \emph{First}, FT1 and FT2 use
the same 100-step full-state checkpoint but different persistence
paths: BeeGFS for FT1 and peer pinned host memory via gloo for
FT2. This tests whether in-memory replication
(Gemini-style~\cite{wang2023gemini}) dominates disk or forms a
Pareto trade-off. \emph{Second}, all remaining \texttt{V}
configurations are included, with FT7 and FT8 reported
analytically. Table~\ref{tab:rq2-mech-frontier} summarizes both.

\subsection{Results}
\label{sec:rq2:results}

\textbf{FT1 (disk) vs FT2 (in-memory) on A2 + F1.}
Both mechanisms checkpoint every 100 steps over a 1500-step
run and persist the same 204.7\,MB state, model parameters plus
SGD momentum buffers post-serialization. BeeGFS scratch
delivers $\sim$1.37\,GB/s write and $\sim$1.32\,GB/s read,
comparable to the gloo-over-IB ring bandwidth FT2 uses, so the
comparison is not confounded by raw transport speed.

The two sit on a genuine Pareto trade-off, not a dominance
relation. FT1 achieves $\sim$8$\times$ lower SS overhead
($0.48\%$ vs $3.78\%$) despite only $\sim$1.5$\times$ lower
per-save cost ($283$\,ms vs $423$\,ms), because their
foreground coupling differs: FT1 writes only from rank 0 with
no barrier, so the other seven ranks overlap the write with
compute, whereas FT2's coordinated ring send/recv puts all
eight ranks on the critical path. FT2 achieves $\sim$9$\times$
lower \ensuremath{T_{\text{restore}}} ($\sim$17\,ms vs
$\sim$155\,ms), restoring from a pinned host buffer in a single
PCIe transfer rather than re-reading 204.7\,MB from BeeGFS. The
$9\times$ is a ratio of analytical bounds; M1's cold-cache
measured values ($20.5$ vs $340.2$\,ms,
Table~\ref{tab:m1-validation}) widen it to $\sim$17$\times$, so
the ordering is robust.
Pareto-incomparability does not by itself settle which to
deploy. Both checkpoint at the same interval and so lose the
same work on a fault, leaving a 3.3-point overhead gap against
a 138\,ms restore gap; scalarizing by expected completion time
therefore favors FT1 unless faults arrive every few seconds. We
make that threshold explicit in the decision framework.

\textbf{FT4 emergency save on A2.}
FT4 inherits Gemini's~\cite{wang2023gemini} peer-pair in-memory
storage but with Bamboo's~\cite{thorpe2023bamboo} reactive
trigger, conceptually closest to the JIT-checkpoint mechanism
of Gupta et al.~\cite{gupta2024jit}: it performs a single save
inside the lead-time window opened by a predictive fault
signal. We use a 2.0\,s lead time. Across three seeds the emergency save completes in
$884.7\pm89.8$\,ms, well inside the window; the dominant cost
is pinned-buffer allocation, which FT2 amortizes at attach time
and FT4 must pay on the critical path. Restore reuses FT2's
peer-pinned-to-GPU path, so FT4 inherits FT2's $\sim$17\,ms
\ensuremath{T_{\text{restore}}}.

\textbf{Multi-mechanism Pareto cross-section on A2.}
Table~\ref{tab:rq2-mech-frontier} places FT4 alongside FT1, FT2,
and the FT9 no-FT baseline on the A2 + F1 axis, and adds
FT3, FT5--FT8 for a complete cross-section.
The \emph{per-save} column is the warm amortized cost for
periodic mechanisms (FT1--FT3, FT5, FT6) and the one-shot
pre-fault cost for FT4. FT4 performs no periodic per-step
checkpoint work, so its SS entry is a design value rather than
a measured throughput difference. The \emph{effective \ensuremath{T_{\text{restore}}}} column adds
the save-at-failure cost to the restore cost for FT4 because
the emergency save is on the critical path, whereas FT1--FT3
and FT5--FT6 amortize theirs off it. The \emph{lost work} column is the
expected re-execution under uniform fault arrival: half the
checkpoint interval for periodic mechanisms and $\sim$0 for
FT4, undefined for FT7/FT8, which keep no checkpoint and
reconfigure rather than restore, and unbounded for FT9;
every reported cell fired its fault after the first completed
save. The FT7/FT8 \ensuremath{T_{\text{restore}}} entries
are analytical rendezvous / reinit bounds; FT8's applies a fixed 700\,ms
rendezvous + NCCL communicator-setup budget identically to
every architecture.

We construct the Pareto frontier over FT1--FT6, whose reported
values represent comparable state-preservation paths. FT7 and
FT8 appear in Table~\ref{tab:rq2-mech-frontier} for
completeness but are excluded from the frontier: recovery
requires external reconfiguration the harness does not execute,
so their latencies are analytical lower bounds rather than
comparable measurements.

\begin{table}[h]
\caption{A2 Ring-AR + ResNet-50 cross-section
(8$\times$V100). For FT2, FT3, FT5 and FT6, steady-state and
save-cost statistics pool two seeds from each of F1--F3
($n\!=\!6$); F1 fixes the recovery operating point. Effective
\ensuremath{T_{\text{restore}}} is labeled \markm{} (measured),
\marka{} (analytical), or \markm$+$\marka{} for FT4
(measured save + analytical restore).}
\label{tab:rq2-mech-frontier}
\centering
\small
\begin{adjustbox}{max width=\columnwidth}
\begin{tabular}{@{}l r r r r r@{}}
\toprule
Configuration & $n$ & SS overhead & Per-save (ms) & Effective & Lost work \\
          &     &             &               & \ensuremath{T_{\text{restore}}} (ms) & (steps) \\
\midrule
FT1 disk (BeeGFS)            & 3 & $0.48\pm0.03\%$          & $283.3$ (warm)        & $\sim$155~\marka & $\sim$50 \\
FT2 in-mem 100-step          & 6 & $3.78\pm0.16\%$          & $423.4\pm17.6$ (warm) & $\sim$17~\marka & $\sim$50 \\
FT3 async-lazy disk          & 6 & $1.19\pm0.04\%$          & $133.7\pm5.0$ (snap)  & $\sim$300~\marka & $\sim$50 \\
\textbf{FT4 JIT 2\,s lead}   & 3 & $0\%$ (design)        & $884.7\pm89.8$        & $\sim$902~\markm$+$\marka & $\sim$0 \\
FT5 cont.\ repl.\ (interval 1) & 6 & $372.30\pm41.16\%$     & $417.0\pm46.1$ (warm) & $\sim$17~\marka & $\sim$0.5 \\
FT6 $k$=2 replica            & 6 & $6.75\pm0.34\%$          & $756.0\pm38.4$ (warm) & $\sim$17~\marka & $\sim$50 \\
FT7 redistribution           & 1 & 0\%        & none                  & $\sim$50~\marka & no ckpt \\
FT8 elastic reconfig         & 1 & 0\%        & none                  & $\sim$710~\marka & no ckpt \\
FT9 no-FT baseline           & 1 & 0\%        & none                  & $\infty$ & $\infty$ \\
\bottomrule
\end{tabular}
\end{adjustbox}
\vspace{-0.3cm}
\end{table}

On that basis the A2 frontier carries three non-dominated
points: FT1 disk (low SS, mid \ensuremath{T_{\text{restore}}}),
FT2 in-mem (mid SS, low \ensuremath{T_{\text{restore}}}), and
FT4 JIT (zero periodic SS work, high \ensuremath{T_{\text{restore}}}). FT4
\emph{does not dominate} FT2 because its
\ensuremath{T_{\text{restore}}} is about $50\times$ higher,
and FT2 \emph{does not dominate} FT4 because FT2 incurs
$3.78\%$ steady-state overhead and FT4 incurs no periodic
checkpoint work. The lost-work
column adds a third consideration the two axes do not capture:
at the 100-step interval and A2's $\sim$112\,ms/step
(Table~\ref{tab:rq1-results}), a periodic mechanism re-executes
$\sim$50\,steps $\approx$ 5.6\,s per fault under uniform
arrival~\cite{young1974first,daly2006higher}, whereas FT4
re-executes almost nothing when the warning fires early enough.
FT4's viability is bounded below by its emergency-save cost: a
lead-time signal shorter than $\sim$885\,ms at this state size
degrades it to no-FT for that fault, so production use needs
either a longer-lead signal or composition with FT2 as a
fallback.

Pareto position alone does not determine deployment choice.
Section~\ref{sec:decision} folds failure rate, lost work and
signal coverage into expected completion time and resolves
ranking.

\textbf{FT5 on A2.}
Realized as FT2 with \texttt{interval\_steps}\,=\,1, FT5
costs $372.30\pm41.16\%$ SS at the same $\sim$17\,ms
\ensuremath{T_{\text{restore}}} as FT2, so it is
Pareto-dominated by FT2: both restore in roughly 17\,ms, while
FT5 pays about $100\times$ the steady-state overhead. Its only advantage is progress loss on
fault: at most one re-executed step against up to 99 for FT2 at
interval 100.

\subsection{From Restore Time to End-to-End Recovery}
\label{sec:rq2-pg-cost}
The times above exclude process-group re-formation. End-to-end
recovery is
$T_{\text{recover}} = T_{\text{restore}} + T_{\text{PG}} +
T_{\text{init}}$, and $T_{\text{PG}}$ is not
measurable on the halting architectures. We therefore
examine $T_{\text{PG}}\!\in\![2,10]$\,s as a sensitivity
range. This scale is consistent with rendezvous and
restart costs: the FT8 estimate in
Table~\ref{tab:rq2-mech-frontier} assigns $\sim$700\,ms to the
rendezvous and NCCL setup alone, while
Bamboo~\cite{thorpe2023bamboo} reports multi-second teardown
and restart. Section~\ref{sec:rq3} provides direct
multi-fault reconvergence measurements at the same order of
magnitude.

Holding the mechanism-specific $T_{\text{init}}$ fixed or
negligible, adding the common $T_{\text{PG}}$ term, identical
for FT1, FT2 and FT4 since the process group is re-formed once
per fault regardless of which mechanism populated the
survivors, preserves the ordering
FT2\,$\prec$\,FT1\,$\prec$\,FT4 and the absolute gaps
(138\,ms and $\sim$747\,ms), but compresses the ratios: the
FT1/FT2 recovery ratio collapses from $\sim$9$\times$ to
$1.01$--$1.07\times$ as $T_{\text{PG}}$ dominates. Once
$T_{\text{PG}}$ dominates, the recovery axis becomes less
discriminating between mechanisms and steady-state cost becomes
decisive: FT4 has no periodic checkpoint work, while FT1 and FT2
incur 0.48\% and 3.78\% measured SS overhead.

\section{RQ3: Sensitivity Across Deployments}
\label{sec:rq3}

\subsection{Motivation}
\label{sec:rq3:motivation}
RQ2 establishes the mechanism trade-off on a fixed A2
deployment. We next ask how far that trade-off moves when
the training architecture, per-rank state size, or fault regime
changes. If any of them reshuffles the ranking, a single-cell
recommendation is unsafe for production.

\subsection{Approach}
\label{sec:rq3:approach}
\begin{figure}[t]
\centering
\includegraphics[width=\columnwidth]{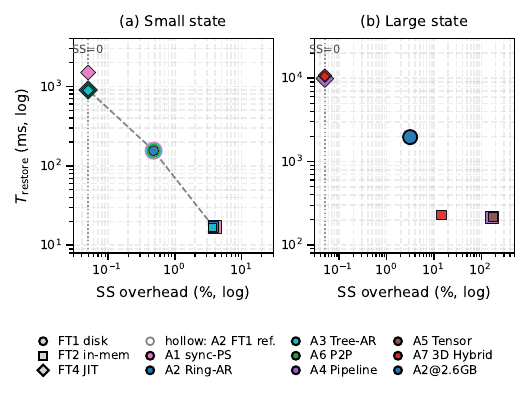}
\caption{Cross-architecture resilience--cost trade-offs under F1.
(a) Small-state architectures ($\sim$205\,MB/rank).
(b) Large-state architectures (2.5--2.7\,GB/rank), including
the matched-state A2 control. Marker shape denotes FT mechanism
and color denotes architecture. Hollow circles mark the A2 FT1
reference, measured only on A2. Zero-overhead FT4 cells are placed
on the dotted floor of the log-scale x-axis.}
\label{fig:rq2-pareto}
\vspace{-0.3cm}
\end{figure}

We evaluate sensitivity along three axes. First, we evaluate
FT2 ($n\!=\!6$ on the measured cells) and FT4
($n\!=\!3$) across the matrix under F1. Second,
M2 reruns A2 with a 2.6\,GB per-rank state to partially
separate state size from architecture. Third, F2 and F3 test
whether the single-fault conclusions persist under multiple
worker losses. Runs use 8$\times$V100; multi-fault cells
use $n\!\leq\!2$ exploratorily, with the cells used by the
decision framework boosted to $n\!=\!5$.
Figure~\ref{fig:rq2-pareto} plots the F1 sweep on the
(SS overhead, \ensuremath{T_{\text{restore}}}) plane.

\subsection{Results}
\label{sec:rq3:results}

\textbf{Architecture sweep: FT2 across the architecture matrix.}
We ran FT2 on every architecture whose FT2 cell is marked
\texttt{V} (viable), with FT2 configured identically (interval =
100~steps, full model + optimizer state, ring replication to the
peer's pinned host memory). Table~\ref{tab:rq2-ft2-cross-arch}
summarises the seven cells. Its A2 Ring-AR (M2) row is the
matched-state control whose deadlock is analysed below
(Table~\ref{tab:m2-deconfound}).

Three findings stand out. (1)~Warm-save throughput is
$\sim$460--510\,MB/s across A2--A5 and A7 despite their
differing topologies and parallelism dimensions; A1 sync-PS sits lower
($\sim$290\,MB/s), consistent with its serialized PS-side
extraction path. This is the gloo-over-IB ring
bandwidth bound: FT2 uses a separate gloo subgroup on
NCCL-primary architectures and the default PG on gloo-primary
ones, so training communication does not contend with it. (2)~SS overhead is the ratio of save cost to step time, so a
nearly constant save cost yields very different percentages
across architectures. A4 pipeline and A5 tensor parallelism exceed
$100\%$ (each FT2 save takes longer than 100 steps of compute
would in their absence) because their per-step compute is fast
(33\,ms) relative to their 2.5\,GB state; A7 hybrid 3D
parallelism sits at $14\%$ only because its per-step compute
is already slow (372\,ms/step). The viable small-state FT2
cells (A1--A3) cluster at $3.69$--$3.91\%$ SS overhead.
(3)~A6+FT2 is structurally infeasible because FT2's periodic
save needs a long-lived full-rank ring process group, which A6
never forms: it keeps only short-lived per-edge groups alive
between gossip rounds. Its 17.17\,ms warm-cache observation is
retained only as a consistency check on the analytical PCIe
restore model, not as an operating point, and FT4's successful
$881.1\pm103.5$\,ms emergency save on A6
(Table~\ref{tab:rq2-ft4-cross-arch}) shows that network
reachability is not the barrier.

\begin{table}[t]
\caption{FT2 cross-architecture characterization on 8$\times$V100
(100-step interval, gloo+IB ring replication to peer pinned host
RAM); mean $\pm$ half-range over two seeds from each of F1--F3
($n\!=\!6$), pooled because steady-state cost is
trace-independent. F1 fixes the recovery point.
A1--A3/A6 use ResNet-50 and A4/A5/A7 Megatron-style MLPs,
so SS overhead reflects both architecture and workload.
A dash denotes an undefined value: A2 (M2) deadlocked before
fault injection, while A6+FT2 is structurally infeasible.}
\label{tab:rq2-ft2-cross-arch}
\centering
\small
\begin{adjustbox}{max width=\columnwidth}
\begin{tabular}{@{}l r r r r r@{}}
\toprule
Arch & State & $n$ & Warm save & SS\,oh. & \ensuremath{T_{\text{restore}}} \\
     & (MB)  &     & (ms)      & (\%)    & (ms) \\
\midrule
A1 sync-PS         & 204.7  & 6 & $714.9\pm24.6$  & $3.91\pm0.13$   & $\sim$17~\marka \\
A2 Ring-AR         & 204.7  & 6 & $423.4\pm17.6$  & $3.78\pm0.16$   & $\sim$17~\marka \\
A2 Ring-AR (M2)    & 2600.0 & 0/3 & deadlock      & --              & -- \\
A3 Tree-AR         & 204.7  & 6 & $424.9\pm17.4$  & $3.69\pm0.15$   & $\sim$17~\marka \\
A4 Pipeline        & 2567.0 & 6 & $5521.9\pm797.2$& $167.33\pm24.16$& $\sim$214~\marka \\
A5 Tensor-Par.\    & 2616.9 & 6 & $5647.7\pm821.5$& $176.49\pm25.67$& $\sim$218~\marka \\
A6 P2P gossip      & 204.7  & 0 & infeasible (\texttt{S}) & --      & 17.17 ($\mu$bench)~\markm \\
A7 3D Hybrid       & 2734.7 & 6 & $5354.8\pm303.8$& $14.39\pm0.82$  & $\sim$228~\marka \\
\bottomrule
\end{tabular}
\end{adjustbox}
\vspace{-0.3cm}
\end{table}

\textbf{Architecture sweep: FT4 across the seven architectures.}
We run FT4 across all seven architectures with lead times
scaled to state size. All seven emergency saves complete within
the warning window (Table~\ref{tab:rq2-ft4-cross-arch}); none
falls back to FT9. Because the save runs only after a warning
arrives, both it and the subsequent restore lie on the critical
path, so the \emph{effective \ensuremath{T_{\text{restore}}}}
column sums the measured emergency save with the analytical
restore cost (state\,MB\,/\,PCIe~Gen-3\,$\sim$12\,GB/s).

Three findings stand out.
(1)~FT4 has zero periodic steady-state checkpoint work by
construction on every architecture: its only per-step hook is
empty.
(2)~Effective \ensuremath{T_{\text{restore}}} ranges from 889\,ms (A3) to nearly 11\,s
(A5, A7), spanning more than a decade, dominated by the
emergency-save cost which scales linearly with state size at the
gloo ring bandwidth ($\sim$500\,MB/s). \emph{FT4's $\sim$0.9\,s small-state
result therefore does not carry over to the 2.5--2.7\,GB
workloads, where its effective \ensuremath{T_{\text{restore}}}
reaches roughly 10\,s.}
(3)~Even at identical state size (204.7\,MB), the
parameter-server architecture (A1) pays a $\sim$70\% higher
emergency-save cost than the other small-state architectures
(A1 $1479.5\pm231.9$\,ms vs.\ A2 $884.7\pm89.8$, A3
$871.8\pm39.3$, A6 $881.1\pm103.5$). The gap is consistent
across all three seeds and consistent with A1's slower state
extraction path, since the PS-only optimizer state is
materialized on rank 0 before serialization begins.

\begin{table}[t]
\caption{FT4 cross-architecture characterization on
8$\times$V100 under F1; values are mean $\pm$ half-range.
Effective \ensuremath{T_{\text{restore}}} combines measured
emergency save with analytical PCIe restore
(\markm$+$\marka{}). A1--A3/A6 use ResNet-50;
A4/A5/A7 use Megatron-style MLPs.}
\label{tab:rq2-ft4-cross-arch}
\centering
\small
\setlength{\tabcolsep}{4pt}
\begin{adjustbox}{max width=\columnwidth}
\begin{tabular}{@{}l r r r r r@{}}
\toprule
Arch & State & $n$ & Lead & Emergency save  & Eff.\,\ensuremath{T_{\text{restore}}} \\
     & (MB)  &     & (s)  & total (ms)      & (ms) \\
\midrule
A1 sync-PS    & 204.7  & 3 & 2  & $1479.5\pm231.9$    & 1{,}497~\markm$+$\marka \\
A2 Ring-AR    & 204.7  & 3 & 2  & $884.7\pm89.8$      & 902~\markm$+$\marka \\
A3 Tree-AR    & 204.7  & 3 & 2  & $871.8\pm39.3$      & 889~\markm$+$\marka \\
A4 Pipeline   & 2567.0 & 3 & 15 & $9697.0\pm330.3$    & 9{,}911~\markm$+$\marka \\
A5 Tensor-Par.& 2616.9 & 3 & 15 & $10558.7\pm1493.4$  & 10{,}777~\markm$+$\marka \\
A6 P2P gossip & 204.7  & 3 & 2  & $881.1\pm103.5$     & 898~\markm$+$\marka \\
A7 3D Hybrid  & 2734.7 & 3 & 15 & $10314.6\pm324.4$   & 10{,}543~\markm$+$\marka \\
\bottomrule
\end{tabular}
\end{adjustbox}
\vspace{-0.3cm}
\end{table}

\textbf{State-size control: A2 with a 2.6\,GB workload.}
The architecture sweep couples state size with architecture
family: A1/A2/A3/A6 train ResNet-50 while A4/A5/A7 train
Megatron-style MLPs. To partially separate them we run A2
Ring-AR with a $\sim$2.6\,GB per-rank state (a single
\texttt{Linear(H, H)} with $H\!=\!18000$, chosen so
$2H^2\cdot 4\,\text{bytes} \approx 2.6\,\text{GB}$), matching
A4/A5/A7's per-rank state. Fault trace F1, $n\!=\!3$ seeds per
mechanism.

Two findings emerge.
\emph{(i)~A larger per-rank state shifts FT1's operating point
with the architecture held fixed.} Despite a
$\sim$13$\times$ larger state and a 5$\times$ shorter
checkpoint interval, FT1's SS overhead rises only from
$0.48\pm0.03\%$ at 205\,MB
(Table~\ref{tab:rq2-mech-frontier}) to $3.14\pm0.24\%$ at
2.6\,GB. Because both cadence and model change, M2 is a
partial rather than a pure state-size deconfound.
\emph{(ii)~FT2 does not scale to 2.6\,GB per-rank state on
A2 as implemented.} All three FT2 seeds terminated on the
NCCL watchdog at 600\,s, timing out on a default-process-group
\texttt{ALLREDUCE} well before the fault fired, so
\emph{training itself deadlocked}. A plausible explanation is
interleaving between FT2's peer-replication subgroup and the
DDP process group at this state size. A4/A5/A7 complete FT2 at the
same state size, consistent with their schedulers streaming
state in smaller shards. This remains a hypothesis, since the
root cause (subgroup participation vs.\ memory-pressure stall)
is not isolated. We therefore report FT2
as not measurable at 2.6\,GB on A2 in its current
implementation.

\textbf{Fault-type sweep: F2 and F3 multi-fault traces.}
Because Python-side \texttt{recover()} does not run on the
halting architectures under fault, we instrument the runner
with a lightweight \emph{recovery supervisor} that watches step
events after each fault and tags each seed \emph{gracedeg}
(training continued), \emph{halted} (no post-fault step event),
or \emph{mixed}. Table~\ref{tab:rq2-f2f3-attribution} reports
each cell's attribution and measured
\ensuremath{T_{\text{recover}}}. Three findings emerge.

(1)~\emph{SS overhead is nearly trace-independent for periodic
mechanisms on A1--A5 and A7.} FT2, FT3, and
FT6 there show at most a 1.1\,pp F1--F2--F3 spread,
consistent with steady-state cost being determined by
per-step work rather than the future fault schedule; FT5
spreads 3--8\,pp against its far larger F1 baseline. A6 is
excluded because its F2/F3 throughput was sampled in the first
30\,s, inside the $\sim$60\,s its per-edge TCP subgroups need
to reach a stable gossip cadence, so we defer to RQ1's A6
measurement (Table~\ref{tab:rq1-results}), which fires the
fault at $t\!=\!180$\,s to escape the transient.

(2)~\emph{Among the tested collective and partitioned
configurations, F2 is not recovered reliably; only A6 gossip
recovers consistently.}
A2$\times$\{FT2,FT3,FT5,FT6\} and A4/A5/A7$\times$FT2 all
halt across seeds. A5+FT4 recovers
on 2/5 seeds ($128\pm2$\,ms) and A7+FT4 on 3/5 seeds
(mean $206$\,s, right-skewed), at $n\!=\!5$ from the M4 seed-boost
campaign. A6 gossip recovers reliably: A6+FT1 F2 at
$2537\pm2272$\,ms ($n\!=\!5$) and A6+FT4 F2 spanning
$\approx$0.1--24\,s across 5 seeds. FT6's $k\!=\!2$
replication was designed to tolerate F2 structurally, but
the runner never activates the second replica
before the collective blocks.

(3)~\emph{F3 cascading is handled structurally on A6 and
halts every non-A6 configuration.} A6's membership service
drops dead ranks in $\sim$1.6\,s; A6+FT1 completes all 2000
steps under F3 on all five M4 seeds, with recovery at
$3905\pm2445$\,ms, $\sim$1.5$\times$ the F2 mean and
consistent with three cascading recoveries whose
reconvergence windows overlap. All tested non-A6 F3 cells halt
within a few hundred steps of the first fault because the
runner cannot re-form the process group at $N{-}2$ ranks
before fault~2 fires 30\,s later. In these cells F3
therefore distinguishes stored-state availability from
end-to-end runtime reconfiguration; only A6 demonstrates
execution.

\begin{table}[t]
\caption{M2 matched-state control: A2 Ring-AR with a
$\sim$2.6\,GB MLP under F1 ($n\!=\!3$). SS overhead is
relative to the same-state FT9 baseline, except FT4, whose
zero entry is a design value;
\ensuremath{T_{\text{restore}}} is an analytical lower bound.
Dash denotes runs that did not reach fault injection.}
\label{tab:m2-deconfound}
\centering \small
\setlength{\tabcolsep}{4pt}
\begin{adjustbox}{max width=\columnwidth}
\begin{tabular}{@{}l r r r@{}}
\toprule
Mechanism            & $n$ & SS overhead              & \ensuremath{T_{\text{restore}}} (ms) \\
\midrule
FT1 disk (BeeGFS) 20-step & 3 & $3.14 \pm 0.24\%$     & \vala{1968} \\
FT2 in-mem 100-step  & 0/3 & NCCL deadlock (600\,s)   & -- \\
FT4 JIT 15\,s lead   & 3   & $0\%$ (design)           & \vala{217} \\
FT9 no-FT baseline   & 3   & $0.00 \pm 0.04\%$        & $\infty$ \\
\bottomrule
\end{tabular}
\end{adjustbox}
\vspace{-0.3cm}
\end{table}

Three reporting conventions apply to
Table~\ref{tab:rq2-f2f3-attribution}. For mixed cells, the
finite \ensuremath{T_{\text{recover}}} summarizes only the
seeds that resumed training. Cells labeled \emph{expl.}\
remain at $n\!\leq\!2$ and are not cited in the decision
framework (Table~\ref{tab:decision}); the rest are boosted to
$n\!=\!5$. The three A6+FT4 cells and A7+FT4 F2 are heavily
right-skewed, so mean\,$\pm$\,half-range would place the lower
bound below zero; we therefore report observed ranges for the
A6 cells and the mean alone for A7+FT4.

\textbf{Synthesis.}
The RQ2 trade-off persists, but its operating point shifts
with deployment. On small-state workloads, signal
coverage and storage tier decide the choice rather than the
shape of the frontier: FT2 trades $3.69$--$3.91\%$ SS overhead
for $\sim$17\,ms restore on A1--A3, while FT4 removes periodic
checkpoint overhead at $\sim$0.9--1.5\,s effective restore. On large-state
workloads FT2 rises to $14$--$176\%$ SS overhead while FT4
reaches $9.9$--$10.8$\,s. M2 shows that a large per-rank state
can shift the operating point even with the architecture fixed,
although workload and state size remain coupled. FT5 and FT6 fail on A6 for the same
structural reason as FT2, accounting for nine A6
ring-replication cells. Multi-fault results add a separate
constraint: preserved state alone does
not ensure recovery when the runtime cannot re-form its
communication topology; among the tested F2/F3 cells, only A6
consistently continues.

\begin{table}[t]
\caption{F2/F3 recovery outcomes on 8$\times$V100, with A6 F1
for reference. \ensuremath{T_{\text{recover}}} is
supervisor-measured (\markm).}
\label{tab:rq2-f2f3-attribution}
\centering
\small
\setlength{\tabcolsep}{4pt}
\begin{adjustbox}{max width=8cm}
\begin{tabular}{@{}l l c l l@{}}
\toprule
Arch & FT & Trace & \ensuremath{T_{\text{recover}}} & Outcome \\
\midrule
A2 Ring-AR      & FT2 & F2 & $\infty$              & halted (expl.) \\
A2 Ring-AR      & FT2 & F3 & $\infty$              & halted (expl.) \\
A2 Ring-AR      & FT3 & F2 & $\infty$              & halted (expl.) \\
A2 Ring-AR      & FT3 & F3 & $\infty$              & halted (expl.) \\
A2 Ring-AR      & FT5 & F2 & $\infty$              & halted (expl.) \\
A2 Ring-AR      & FT5 & F3 & $\infty$              & halted (expl.) \\
A2 Ring-AR      & FT6 & F2 & $\infty$              & halted (expl.) \\
A2 Ring-AR      & FT6 & F3 & $\infty$              & halted (expl.) \\
A4 Pipeline     & FT2 & F2 & $\infty$              & halted (expl.) \\
A4 Pipeline     & FT4 & F2 & $\infty$              & halted (expl.) \\
A5 Tensor-Par.  & FT2 & F2 & $\infty$              & halted (expl.) \\
A5 Tensor-Par.  & FT4 & F2 & $128\pm2$\,ms                & mixed (2/5 grace) \\
A6 P2P gossip   & FT1 & F1 & $18367\pm12364$\,ms & gracedeg (expl.) \\
A6 P2P gossip   & FT1 & F2 & $2537\pm2272$\,ms          & gracedeg \\
A6 P2P gossip   & FT1 & F3 & $3905\pm2445$\,ms          & gracedeg \\
A6 P2P gossip   & FT4 & F1 & range $\approx$0.1--23\,s & gracedeg \\
A6 P2P gossip   & FT4 & F2 & range $\approx$0.1--24\,s & gracedeg \\
A6 P2P gossip   & FT4 & F3 & range $\approx$3--33\,s   & gracedeg \\
A7 3D Hybrid    & FT2 & F2 & $\infty$              & halted (expl.) \\
A7 3D Hybrid    & FT4 & F2 & $206$\,s (mean, skewed) & mixed (3/5 grace) \\
\bottomrule
\end{tabular}
\end{adjustbox}
\vspace{-0.3cm}
\end{table}
\section{Decision Framework}
\label{sec:decision}

\begin{table*}[t]
\caption{Decision framework from Eq.~\eqref{eq:cost} using
measured and extrapolated operating points. Selections are
among characterized candidates; evidence and latency semantics
are defined in the text.}
\label{tab:decision}
\centering
\small
\setlength{\tabcolsep}{3pt}
\begin{adjustbox}{max width=18cm}
\begin{tabular}{@{}l l l l l r r c@{}}
\toprule
Architecture class & State per rank & Failure regime & Binding constraint
& Selected FT & SS oh. & Relevant latency & Ev. \\
\midrule
Small-state DDP/PS/AR & $\sim$205\,MB & F1, signaled
  & Warning lead\,$\geq$\,2\,s available & \textbf{FT4}, lead\,=\,2\,s
  & 0\% (design) & $0.89$--$1.50$\,s & M \\
Small-state DDP/PS/AR & $\sim$205\,MB & F1, unsignaled
  & Fast tier ($\geq$\,1\,GB/s), no restore cap & \textbf{FT1}, interval\,=\,100
  & $0.48\%$ & $\sim$155\,ms & E \\
Small-state DDP/PS/AR & $\sim$205\,MB & F1, unsignaled
  & Slow tier, or restore cap $<$\,150\,ms & \textbf{FT2}, interval\,=\,100
  & $3.7$--$3.9\%$ & $\sim$17\,ms & M \\
Large-state DDP/AR (M2) & $\sim$2.6\,GB & F1, unsignaled
  & Continuous oh.\ tolerated up to 5\% & \textbf{FT1}, interval\,=\,20
  & $3.14\%$ & $\sim$1.97\,s & M \\
Large-state pipeline / TP & 2.5--2.7\,GB & F1, unsignaled
  & Continuous oh.\ tolerated up to 35\% & \textbf{FT2}, interval\,=\,500
  & $\sim$33\% & $\sim$0.2\,s & E \\
Large-state pipeline / TP & 2.5--2.7\,GB & F1, signaled
  & Warning lead\,$\geq$\,15\,s available & \textbf{FT4}, lead\,=\,15\,s
  & 0\% (design) & $9.9$--$10.8$\,s & M \\
Large-state 3D hybrid     & 2.73\,GB    & F1, unsignaled
  & SS-tolerated ($\leq$\,15\%) & \textbf{FT2}, interval\,=\,100
  & 14\%     & $\sim$0.23\,s & M \\
Gossip/D-PSGD (A6)        & $\sim$205\,MB & F1 single fail-stop
  & Graceful degradation already provides FT
  & \textbf{none} (FT9)\,$+$\,membership svc.\
  & 0\%      & $\sim$0.5--1.6\,s & D \\
Async-PS (A1 async)       & $\sim$205\,MB & F1 single fail-stop
  & Graceful degradation already provides FT
  & \textbf{none} (FT9)\,$+$\,receive timeout
  & 0\%      & $\sim$5\,s & D \\
Gossip / D-PSGD (A6)      & $\sim$205\,MB & F2 or F3 multi-fault
  & Any (arch handles it)
  & \textbf{none} (FT9)\,$+$\,membership svc.\
  & 0\%      & 2.5--3.9\,s & E \\
Tested non-A6 configs.\ & Any    & F2 or F3 multi-fault
  & Any
  & Compose FT with external election
  & Mech.\ dep. & Runner-side & -- \\
\bottomrule
\end{tabular}
\end{adjustbox}
\vspace{-0.4cm}
\end{table*}

\emph{What ``cost'' means.} We minimize expected job completion
time; storage, host memory, and device memory enter only as
feasibility constraints, since a mechanism whose replica does not fit
is not a candidate. For no-FT failure-free wall time $T_0$ and
mechanism $m$ of steady-state overhead $O_m$, with fail-stop events
arriving at mean time between failures $M$,
\begin{equation}
\mathbb{E}[T_{\text{tot}}] = T_0(1{+}O_m)
  + \mathbb{E}[N_f]\bigl(T_{\text{det}} + T_{\text{rec}}
  + T_{\text{lost}}\bigr),
\label{eq:cost}
\end{equation}
where $\mathbb{E}[N_f] \approx T_0(1{+}O_m)/M$, neglecting the
second-order chance of a fault arriving during recovery. Every
coefficient is measured, bounded, or derived from the empirical
tables. $T_{\text{det}}$ is detection latency, $T_{\text{rec}} =
T_{\text{restore}} + T_{\text{PG}} + T_{\text{init}}$, and
$T_{\text{lost}} = \tfrac{1}{2}I\,t_{\text{step}}$ for a periodic
mechanism at interval $I$, near zero for a reactive one under a
covered warning. Across mechanisms needing the same process-group
reconstruction on a fixed architecture, $T_{\text{PG}}$ is common and
shifts their recovery costs equally. Rows based on native graceful
degradation fall outside this additive model, continuing at a changed
post-fault operating point, and are reported directly in
Table~\ref{tab:decision}. Selection minimizes Eq.~\eqref{eq:cost}
over the characterized viable candidates subject to a steady-state
cap $O_m \leq c_{\text{ss}}$ and a pause-on-fault cap. On the halting
architectures $T_{\text{det}}$ is common to the compared candidates,
so Table~\ref{tab:decision} states the actionable restore cap
$T_{\text{restore}} \leq c_{\text{restore}}$.

\emph{Which term dominates.} The overhead term scales with $T_0$ and
the recovery term with $T_0/M$, so of two mechanisms differing by
$\Delta O$ in overhead and $\Delta C$ in per-fault cost, the
higher-overhead one wins only when $M < \Delta C / \Delta O$. On A2 at
205\,MB, FT1 and FT2 checkpoint at the same interval and so lose the
same work, leaving $\Delta O = 3.3$ points against $\Delta C =
138$\,ms: FT2 minimizes completion time only below $M \approx
4.2$\,s, or $9.7$\,s using the cold-cache measured restores. Both are
far below the production failure intervals reported
in~\cite{zhang2022opt,llama3card,jiang2024megascale}, so scalarizing
by completion time orders a pair RQ2 reports as Pareto-incomparable:
FT1 minimizes expected time unless the restore cap binds or the
storage tier is slow enough to lift its overhead above the cap. FT4
is gated by signal coverage instead. Under a warning whose lead time
exceeds the emergency-save cost it carries $O = 0$ by design and
$T_{\text{lost}} \approx 0$, which makes it the Eq.~\eqref{eq:cost}
minimizer over the measured A2 candidates at every $M$; on the
large-state cells its $\sim$10\,s effective restore must be weighed
against FT2's restore under 230\,ms. Without a covered warning it
preserves no state and is not a candidate at all.

Table~\ref{tab:decision} instantiates the framework from these
measurements and stated extrapolations. Rows with mechanism
selections carry an evidence code: \textbf{M} measured directly,
\textbf{E} extrapolated under a stated assumption, and \textbf{D}
derived from the architecture's own no-FT degradation, where the
measured answer is that no mechanism is needed for crash-only faults.
``Relevant latency'' means effective
\ensuremath{T_{\text{restore}}} for the state-preserving mechanisms
and measured detection or reconvergence latency for the
graceful-degradation rows. Selections are among the characterized
candidates rather than the full viable set, since FT1, for instance,
is not measured on A4/A5/A7. Absolute values should be re-derived on
the target cluster; the transferable axes are per-rank state size and
signal coverage.

\emph{Reading the table.} Match on architecture class and state, then
on failure regime and binding constraint. The three small-state rows
are alternative regimes rather than competing answers: a covered
warning selects FT4, and its absence hands the choice to the storage
tier and the restore cap. The small-state FT1 row is coded \textbf{E}
because its $0.48\%$ and 155\,ms are measured on A2 only, whereas the
FT4 row is \textbf{M} across all four small-state architectures, its
range spanning A3 (889\,ms) to A1 (1497\,ms). The M2 FT1 row is
measured, and FT2 is excluded there because all three M2 seeds
deadlocked before fault injection. The subsequent
pipeline\,/\,TP FT2 row is \textbf{E},
projected from the 100-step cell of
Table~\ref{tab:rq2-ft2-cross-arch} on the assumption that SS overhead
scales inversely with interval. The A6 F2/F3 row is \textbf{E}
because its latency comes from runs with FT1 or FT4 attached but is
dominated by native membership handling; its spread is wide enough
that a deployment should provision for several seconds of
reconvergence rather than the mean alone. The final row is a
requirement rather than an Eq.~\eqref{eq:cost} minimizer: no
mechanism we evaluated completes end-to-end recovery under F2 or F3
without external orchestration. Relatedly, on the halting
architectures we verify \emph{recoverability} but not completed
recovery, so $T_{\text{rec}}$ is a lower bound there.

\section{Threats to Validity}
\label{sec:threats}

\textbf{Internal validity.}
Injector timing skew is below 20\,ms, and we discard the first
50 steps of each run. Sub-20\,ms analytical restore bounds are
not treated as measured end-to-end recovery. The
A1-sync/A1-async comparison changes both backend and
synchronization policy, and A1-async uses responder threshold
one, so the server proceeds on a single reply and its graceful
degradation under one fault is close to guaranteed by
construction; its result therefore characterizes this
configuration rather than asynchronous parameter servers
generally. On NCCL-primary architectures,
\ensuremath{T_{\text{restore}}} is an analytical lower bound;
Table~\ref{tab:m1-validation} shows measured costs 20--120\%
higher. Fault victims are fixed by trace, limiting
generalization of position-dependent effects.

\textbf{Construct and recovery validity.}
SS overhead measures amortized FT cost relative to the
failure-free baseline. FT4's zero SS entry is structural, not a
throughput estimate: it does no periodic checkpoint work, and its
emergency save is charged to effective
\ensuremath{T_{\text{restore}}}. FT7/FT8 report only pre-fault SS
overhead, having no state-preservation path. On
the halting architectures (A1-sync, A2--A5, A7), we establish
recoverability but not completed end-to-end recovery because
the runner does not perform survivor election and process-group
reconstruction. A6 reconfigures natively, while A1-async is
native graceful continuation rather than FT recovery.

\textbf{Coverage and implementation fidelity.}
Exploratory F2/F3 cells use $n\!\leq\!2$, while cells used by
the decision framework are boosted to $n\!=\!5$; multi-fault
conclusions are therefore limited to the tested configurations.
FT4--FT6 are simplified production proxies, and A5 omits
Megatron's full autograd wrappers. They preserve the
state-transfer and failure structures targeted by our metrics
but are not end-to-end convergence evaluations. For
small-$n$ FT experiments, we report mean $\pm$ half-range.

\textbf{External validity.}
Measurements use 8 V100 GPUs across two nodes with HDR
InfiniBand and BeeGFS scratch ($\sim$1.37\,GB/s write,
$\sim$1.32\,GB/s read), whereas the cited production runs
operate at much larger scale, so absolute overheads and
crossover points may change with filesystem, interconnect, and
world size. We use ResNet-50 and Megatron-style feed-forward
stacks rather than a full Transformer; the workloads preserve
key determinants of our metrics, including per-rank state size
and communication structure, but activation checkpointing,
larger Adam-style optimizer state, and sequence-parallel
communication remain outside our measurements. Structural
results are less timing-dependent, but crossover points should
be re-derived on the target system.

\textbf{A6 learning efficiency.}
A6's post-fault throughput increase does not imply improved
optimization. Truncating both groups to a 1123\,s window
from the fault instant, neither loss reduction nor time to
target differs detectably against the matched no-fault baseline
($p\!=\!0.67$ and $p\!=\!0.83$; $n\!=\!7$ fault, $n\!=\!3$
baseline), despite roughly 18\% higher sample throughput.
Validation on a real training workload remains open.

\textbf{Fault-model scope.}
F1--F3 inject fail-stop worker crashes and never fail the A1
parameter-server coordinator. Coordinator failure, stragglers,
and network partitions are outside scope; they interact
differently with architectures and FT mechanisms and require
dedicated fault traces.

\section{Conclusion}
\label{sec:conclusion}

We presented \sysname{}, a unified evaluation of crash fault
tolerance across seven distributed-training architectures,
eight FT mechanisms, and a no-FT baseline, covering 142
(architecture, mechanism, trace) tuples.

Three findings stand out. First, six configurations halt after
a single worker failure, whereas A1-async and A6 gossip
degrade gracefully; A6's post-fault sample throughput rises
17.9\% without a detectable improvement in loss progress.
Second, disk, in-memory, and JIT checkpointing form a Pareto
frontier on A2, but their operating points shift strongly with
architecture and state size. Third, compatibility itself limits
mechanism transfer: some architecture--mechanism pairs are
structurally infeasible rather than merely unevaluated.

We translate these results into a measurement-grounded
decision framework that selects among characterized candidates
using failure rate, state size, checkpoint interval, signal
availability, and service-level constraints. Future work should
validate these trade-offs on larger clusters, full Transformer
workloads, and broader fault models.

\bibliographystyle{IEEEtran}
\bibliography{references}


\begin{IEEEbiography}[{\includegraphics[width=1in,height=1.25in,%
  clip,keepaspectratio]{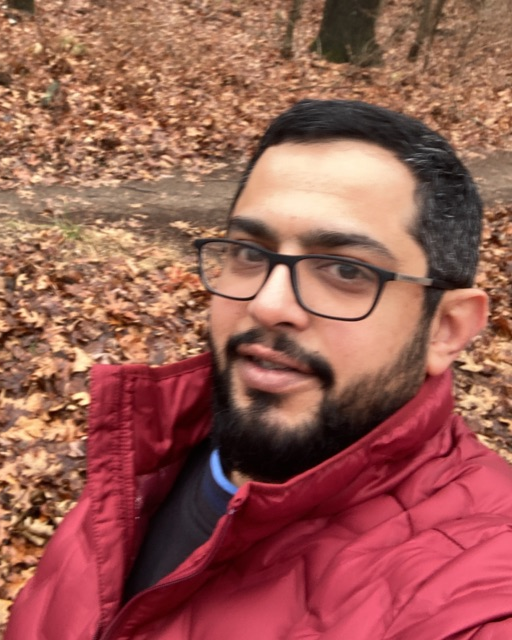}}]{Khaled Aljbab}
received the B.Sc. degree in computer science from Jouf
University and the M.S. degree in computer science from
California Lutheran University.
He is currently pursuing the Ph.D. degree in computer science
and engineering at Oakland University. His research interests
include fault tolerance and fair scheduling for distributed
machine learning systems.
\end{IEEEbiography}

\begin{IEEEbiography}[{\includegraphics[width=1in,height=1.25in,%
  clip,keepaspectratio]{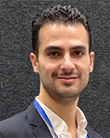}}]{Amine Barrak}
received the M.Sc. degree in software engineering from
Polytechnique Montr\'eal and the Ph.D. degree in computer
science from the Universit\'e du Qu\'ebec. He is currently an Assistant Professor with the Department of Computer Science and Engineering, Oakland University. His
research interests include MLOps, serverless computing,
distributed and federated machine learning, and cloud-native
systems for large-scale AI.
He received the Tunisian Government Excellence
Scholarship and the Best Student Paper Award at CASCON2019.
\end{IEEEbiography}

\end{document}